\documentclass[10pt]{iopart}

\usepackage{graphicx}
\usepackage{dcolumn}
\usepackage{bm}
\usepackage{color}

\begin{document}

\title[AE saturation]{Effect of the neutral beam injector operational regime on the Alfven eigenmode saturation phase in DIII-D plasma}


\author{J. Varela}
\ead{jvrodrig@fis.uc3m.es}
\address{Universidad Carlos III de Madrid, 28911 Leganes, Madrid, Spain}
\author{D. A. Spong}
\address{Oak Ridge National Laboratory, Oak Ridge, Tennessee 37831-8071, USA}
\author{L. Garcia}
\address{Universidad Carlos III de Madrid, 28911 Leganes, Madrid, Spain}
\author{Y. Ghai}
\address{Oak Ridge National Laboratory, Oak Ridge, Tennessee 37831-8071, USA}
\author{D. Zarzoso}
\address{Aix-Marseille Université, CNRS, Centrale Marseille, M2P2 UMR 7340, Marseille, France}
\author{D. del-Castillo-Negrete}
\address{Oak Ridge National Laboratory, Oak Ridge, Tennessee 37831-8071, USA}
\author{H. Betar}
\address{Aix-Marseille Université, CNRS, Centrale Marseille, M2P2 UMR 7340, Marseille, France}
\author{J. Ortiz}
\address{Universidad Carlos III de Madrid, 28911 Leganes, Madrid, Spain}
\author{D. C. Pace}
\address{General Atomics, P.O. Box 85608, San Diego, California 92186-5608, USA}
\author{M. A. Van Zeeland}
\address{General Atomics, P.O. Box 85608, San Diego, California 92186-5608, USA}
\author{X. Du}
\address{General Atomics, P.O. Box 85608, San Diego, California 92186-5608, USA}
\author{R. Sanchez}
\address{Universidad Carlos III de Madrid, 28911 Leganes, Madrid, Spain}
\author{V. Tribaldos}
\address{Universidad Carlos III de Madrid, 28911 Leganes, Madrid, Spain}
\author{J. M. Reynolds-Barredo}
\address{Universidad Carlos III de Madrid, 28911 Leganes, Madrid, Spain}

\date{\today}

\begin{abstract}
The aim of this study is to analyze the effect of the neutral beam injector (NBI) operation regime on the saturation phase of the Alfven Eigenmodes (AE) in DIII-D plasma. The analysis is done using the linear and nonlinear versions of the gyro-fluid code FAR3d. A set of parametric analyses are performed modifying the nonlinear simulation EP $\beta$ (NBI injection power), EP energy (NBI voltage) and the radial location of the EP density profile gradient (NBI radial deposition). The analysis indicates a transition from the soft (local plasma relaxation) to the hard MHD (global plasma relaxation) limit if the simulation EP $\beta \geq 0.02$, leading to bursting MHD activity caused by radial AEs overlapping. MHD bursts cause an enhancement of the EP transport showing ballistic-like features as avalanche-like events. Simulations in the soft MHD limit show an increment of the EP density gradient as the EP $\beta$ increases. On the other hand, there is a gradient upper limit in the hard MHD limit, consistent with the critical-gradient behavior. AEs induce shear flows and zonal current leading to the deformation of the flux surfaces and the safety factor profile, respectively, particularly strong for the simulation in the hard MHD limit. Simulations in the hard MHD regime show a decrease of the AE frequency in the saturation phase; this is caused by the destabilization of a transitional mode between a $9/3-10/3$ TAE and a $9/3$ RSAE that may explain the AE frequency down-sweeping observed in some DIII-D discharges. Reducing the EP energy in the nonlinear simulations leads to a weakening of the plasma perturbation. On the other hand, increasing the EP energy causes the opposite effect. Nonlinear simulations of off-axis NBI profiles indicate a lower plasma perturbation as the EP density gradient is located further away from the magnetic axis.
\end{abstract}

%
%
%
%
%


\vspace{2pc}

\maketitle

\ioptwocol

\section{Introduction \label{sec:introduction}}

The efficient plasma heating of future nuclear fusion reactors requires avoiding or minimizing the  activity of energetic particle (EP) driven instabilities. EP driven instabilities enhance the transport of neutral beam injector (NBI) and ion cyclotron resonance heating (ICRH) EPs as well as alpha particles \cite{1,2,3,4,5,6,7}. Experiments in several Tokamaks and Stellarators show a degradation of the device performance by EP driven instabilities leading to larger EP diffusive losses \cite{8,9,10,11,12,13,14}.

EPs and alpha particles can destabilize Alfv\' en Eigenmodes (AE) if their transit, bounce or drift frequencies resonate with AE mode frequencies \cite{15,16}, triggered in the spectral gaps of the shear Alfv\' en continua \cite{17,18}. In addition, if the EP drive is strong enough to exceed the continuum damping effect, energetic particle modes (EPMs) can be triggered inside the continuum \cite{19,20}.

The DIII-D plasma is heated by eight NBI ($2.25$ MW / source of power) that inject deuterium with a beam energy of $80$ keV in a deuterium plasma, inducing a rich AE / EPM activity \cite{21,22,23,24}. The variation of the Alfv\' en speed lead to the destabilization of different AE families: toroidicity induced Alfv\' en Eigenmodes (TAE) coupling $m$ with $m+1$ modes (m is the poloidal mode number) \cite{25,26}, ellipticity induced Alfv\' en Eigenmodes (EAE) coupling $m$ with $m+2$ modes \cite{27,28} or noncircularity induced Alfv\' en Eigenmodes (NAE) coupling $m$ with $m+3$ or higher \cite{29,30}. In addition, $\beta$ induced Alfv\' en Eigenmodes (BAE) are driven by compressibility effects \cite{31}, Reversed-shear Alfv\' en Eigenmodes (RSAE) at the local maxima/minima in the safety factor $q$ profile \cite{32,33} and Global Alfv\' en Eigenmodes (GAE) nearby the minimum of the Alfv\' en continua \cite{34,35}. DIII-D experiments measured an enhancement of the transport and EP losses linked to unstable AE \cite{36,37,38,39}.

Several techniques were tested in DIII-D plasma to improve the AE stability. For example, the application of external actuators as the electron cyclotron heating (ECH) \cite{40,41,42} and electron cyclotron current drive (ECCD) \cite{43,44} or the optimization of the NBI operational regime \cite{45,46,47}. Several methods were explored in other devices such as the application of ICRH in JET plasma \cite{48,49}, NBI current drive in LHD \cite{50} and resonant magnetic perturbations (RMPs) in NSTX \cite{51}. 

The combination of experiments and numerical models provides useful information to identify operation scenarios with an improved AE/EPM stability and plasma heating efficiency \cite{50,52,53,54,55,56,57,58,59,60,61}. Nevertheless, optimization analyses are based on the linear stability of AE/EPM and thus no information is provided about their saturation phase \cite{62,63,64}. 

The aim of this study is to analyze the effect of the NBI operational regime on the AE saturation phase in DIII-D plasma, particularly the NBI injection power (simulation EP $\beta$), voltage (EP energy) and deposition region (EP density profile). A set of nonlinear simulations are performed identifying the effects of the non linear mode coupling, energy transfers to the thermal plasma, evolution of the equilibrium profiles and the generation of shear flows / zonal currents. The simulations are performed by the gyro-fluid code FAR3d that solves the reduced linear or nonlinear resistive MHD equations coupled with equations of the EP density and parallel velocity \cite{65,66,67,68}. The Landau damping/growth is included in the model through Landau closure relations, thus the model includes the linear wave-particle resonance effects on six field variables that evolve from a three dimensional equilibria generated by the VMEC code \cite{69}.  

This paper is organized as follows. An introduction to the numerical scheme is presented in section \ref{sec:model}. The analysis of the AE linear stability is performed in section \ref{sec:linear}. Next, the effect of the NBI power injection, voltage and deposition region on the AE saturation phase is analyzed in section \ref{sec:nonlinear}. The effect of the equilibrium profiles evolution on the AE linear stability and implications on the critical-gradient paradigm are studied in section \ref{sec:linear2}. Finally, the conclusions of this paper are presented in section \ref{sec:conclusions}.

\section{Equations and numerical scheme \label{sec:model}}

The gyro-fluid FAR3d code solves the linear and nonlinear reduced resistive MHD equations describing the thermal plasma evolution coupled with the first two moments of the gyro-kinetic equation, the equations of the EP density and parallel velocity moments \cite{65,70}, introducing the wave-particle resonance effects required for Landau damping/growth. In addition, the model includes the parallel momentum response of the thermal plasma required for coupling to the geodesic acoustic waves \cite{71}. The code variables evolve in an equilibria calculated by the VMEC code \cite{69} based on conversions of EFIT equilibria. The model calibration is based on the Landau closure coefficients that are calculated from the response function obtained from gyro-kinetic simulations. The closure coefficients are selected to match analytic Toroidal AE (TAE) growth rates based upon a two-pole approximation of the plasma dispersion function with a Lorentzian energy distribution function for the EP. The lowest order Lorentzian can be matched either to a Maxwellian or to a slowing-down distribution by choosing an equivalent average energy. Further details of the model equations can be found in the references \cite{62,72}. The time evolution of the EP energy is not included in this version of the code. Thus, the properties of the EP resonance with respect to the EP energy remains fixed along the simulation. This model simplification is equivalent to assume there is a dominant EP resonance that directs the main part of the available free energy of the system to trigger AEs, neglecting secondary resonances.

FAR3d simulations were validated in benchmarking studies compared with the output of gyro-kinetic and hybrid codes \cite{73}. FAR3d was also used used to analyze the MHD activity in LHD \cite{23,55,68,74,75}, DIII-D \cite{46,52,58,60,76}, TJ-II \cite{59,77,78,79} and Heliotron J \cite{56,57,61} plasmas, showing a reasonable agreement between simulations and experimental data. In addition, the nonlinear version of the code was applied in the analysis of sawtooth-like events, internal collapse events, EIC burst and MHD burst observed in LHD plasma \cite{63,64,80,81,82,83} as well as the AE saturation in DIII-D plasma \cite{62}. The AE stability in plasma with multiple EP species for LHD, DIII-D, ITER and CFETR plasma \cite{54,84,85} as well as predictions of the AE stability in future devices as JT60SA \cite{86} and new devices exploiting different symmetries of the magnetic trap as CFQS and QPS \cite{87,88} were also analyzed using the FAR3d code.

\subsection{Equilibrium properties}

A fixed boundary equilibrium is calculated by the VMEC code based on the DIII-D discharge $159243$ during L-mode current ramp phase at $t = 805$ ms \cite{73}. The experimental observations indicate the destabilization of multiple RSAEs and TAEs while a combination of beams are injected. The equilibrium profiles are obtained from kinetic EFIT calculations constrained by Motional Stark Effect measurements, external magnetics data and q profile information derived from the AE activity. The EP profiles are calculated using the kick model and the TRANSP-NUBEAM code \cite{89,90,91,92}.

\begin{figure}[h!]
\centering
\includegraphics[width=0.5\textwidth]{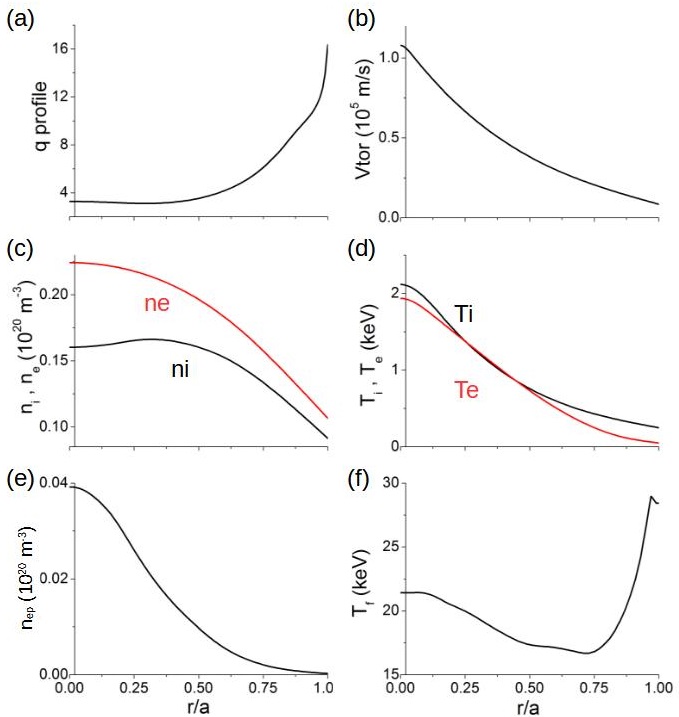}
\caption{Main profiles of the mode. (a) q profile, (b) equilibrium toroidal plasma rotation, (c) thermal electron (red line) and ion (black line) density, (d) thermal electron (red line) and ion (black line) temperature, (e) EP density and (f) EP energy.}\label{FIG:1}
\end{figure}

Figure~\ref{FIG:1} indicates the main profiles of the model. The panel (a) shows the q profile, panel (b) the equilibrium toroidal plasma rotation, panel (c) the thermal electron and ion density, panel (d) the thermal electron and ion temperature, panel (e) the EP density and panel (f) the EP energy. The q profile shows a weak reverse shear region around $r/a = 0.3$ and the strongest gradient of the EP density is observed in the inner plasma region.

The birth energy of the particles injected by the NBI is $E_{inj} = 70$ keV. The model assumes a nominal temperature of the passing EP of $T_{f}(0) = 21.4$ keV.

\subsection{Simulations parameters}

The dynamic toroidal modes ($n$) included in the linear simulations are $n=1$ to $6$ and the dynamic poloidal modes ($m$) are selected to cover the resonant modes in the plasma from the magnetic axis to the plasma periphery ($r/a > 0.8$). The number of toroidal families in the nonlinear simulations is reduced to the $n=3$ and $6$ modes. The identification of the mode number is consistent with the definition of the q profile, $q=m/n$.

\begin{table}[t]
\centering
\begin{tabular}{c c c}
\hline
n & m (linear) & m (nonlinear) \\ \hline
0 & [0,14] & [0,14]\\
1 & [2,10] & NONE \\
2 & [2,20] & NONE \\
3 & [3,25] & [1,20] \\
4 & [4,30] & NONE \\
5 & [5,35] & NONE \\
6 & [6,40] & [18,52] \\ \hline
\end{tabular}
\caption{Equilibrium and dynamic modes in the linear and nonlinear simulations. The first column indicates the toroidal modes, the second columns the poloidal modes in the linear simulations and the third column the poloidal modes in the nonlinear simulations.} \label{Table:1}
\end{table}

The simulations include both mode parities for all the dynamic variables because the moments of the gyro-kinetic equation break the MHD symmetry. The Fourier decomposition convention used is (pressure example): $n > 0$ indicates the $cos(m\theta + n\zeta)$ and $n<0$ the $sin(m\theta + n\zeta)$. The  eigenfunctions (f) are represented with respect to the sine and cosine components:
\begin{eqnarray} 
f(\rho,\theta, \zeta, t) = \sum_{m,n} f^{s}_{mn}(\rho, t) sin(m \theta + n \zeta) \nonumber\\
+ \sum_{m,n} f^{c}_{mn}(\rho, t) cos(m \theta + n \zeta)
\end{eqnarray}
The EP $\beta$ of the simulation provides the normalization of the EP density profile and it is fixed. The effect of the EP transport induced by the AE is reproduced in the model by the evolution of the EP density profile during the simulation and the drive of the EP is fixed by keeping the EP $\beta$ constant. The effect of sources or sinks is not included in the simulations although, due to the short simulation time required to analyze the AE saturation with respect to the source / sink time scale, the consequences for this analysis are small. Longer simulations reproducing steady state regimes will require introducing the effect of sources / sink. The magnetic Lundquist number is assumed $S=10^5$. The number of radial points is $1000$ for the linear simulations and $400$ for the nonlinear simulations. The simulations include a diffusivity of $D_{i} = 10^{-5}$ in all the variables (normalized to the Alfven time at the magnetic axis ($\tau_{A0}$) and the minor radius (a)). The effect of the Finite Larmor Radius and electron-ion Landau damping are not included here for simplicity. The consequence is an overestimation of the EP drive in the simulations compared to the experiment. Introducing these damping effects is mandatory to reproduce the AE stability measured in the discharge, although the aim of the study is to identify AE stability trends thus FLR and e-i Landau damping effects can be neglected in the first approximation. 

\section{Linear AE stability analysis \label{sec:linear}}

This section is dedicated to study the linear stability of the AEs. The analysis provides information of the linear EP $\beta$ threshold for the AE destabilization, frequency range, radial location, dominant modes and eigenfunction structure of the dominant perturbation for each toroidal mode family. 

\begin{figure}[h!]
\centering
\includegraphics[width=0.5\textwidth]{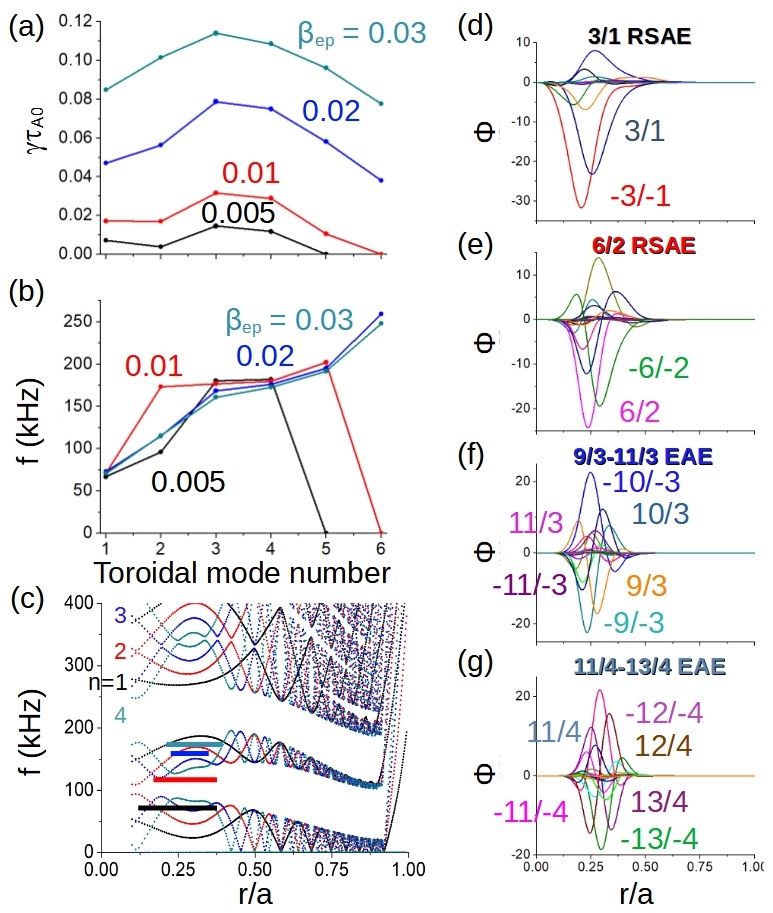}
\caption{Linear AE stability analysis. (a) Growth rate and (b) frequency of the dominant $n=1$ to $6$ AEs. (c) Continuum gaps including the radial location and frequency range of the $n=1$ to $4$ dominant AEs. Eigenfunction of the (d) $3/1$ RSAE, (e) $6/2$ RSAE, (f) $9/3-11/3$ EAE and (g) $11/4-13/4$ EAE.}\label{FIG:2}
\end{figure}

Figure~\ref{FIG:2}, panels a and b, shows the growth rate and frequency of the $n=1$ to $6$ dominant modes for simulations with different EP $\beta$ values. The AE showing the largest growth rate is triggered by the $n=3$ toroidal mode family for all the EP $\beta$ values analyzed. The EP $\beta$ threshold of the $n=1$ to $4$ AEs is $< 0.005$, $0.01$ for the $n=5$ AE and $0.02$ for the $n=6$ AE. Panel c indicates four separate continua, color coded for $n = 1$ (black), $2$ (red), $3$ (blue), and $4$ (cyan) modes as calculated by Stellgap code \cite{93} as well as the radial location and frequency range of the dominant AE triggered in the simulations with EP $\beta = 0.03$. The compressibility index ($\Gamma$) used in Stellgap and FAR3d calculations is $0.02$. The compressibility index is lower than the real plasma to avoid an artificial AE frequency up-shift observed in the simulations with $\Gamma = 5/3$ (for a further description of the approximation consequences please see the references \cite{62,73}). The $n=1$ and $2$ AEs are triggered nearby the q profile minimum and do not intersect with the continuum (black and red lines, respectively), thus these modes are identified as reverse shear AEs. The $n=3$ and $4$ AEs show couplings between $m$ and $m+2$ poloidal modes, identified as EAEs. Panels d to g show the eigenfunctions of the dominant $n=1$ to $4$ AEs if EP $\beta = 0.03$, a $3/1$ RSAE with $f=71$ kHz, a $6/2$ RSAE with $f = 115$ kHz, a $9/3-11/3$ EAE with $f=160$ kHz and a $11/4-13/4$ EAE with $f=173$ kHz, all triggered in the inner plasma region.

The number of toroidal mode families in the nonlinear simulations is reduced to the $n=3$ and $6$. The $n=3$ modes are selected because $n=3$ AEs are the fastest growing modes. On the other hand, $n=6$ is chosen due to the rather large EP $\beta$ threshold for the AE destabilization, that is to say, the simulations with EP $\beta \leq 0.01$ only induce marginally unstable $n=6$ AEs although the cases with EP $\beta \geq 0.02$ leads to the destabilization of robust $n=6$ AEs. Consequently, simulations with EP $\beta \geq 0.02$ represent NBI operational regimes leading to configurations with multiple resonance overlapping, that is to say, a plasma in the hard MHD limit that may show global relaxation events. Simulations with EP $\beta \leq 0.01$ describes a plasma in the soft MHD limit that only induces local relaxations. It should be noted that the simulations require introducing at least $n=1$ to $6$ modes to perform a realistic description of the plasma stability in the discharge $159243$ because unstable AEs induced by all these toroidal mode families were measured during the experiment \cite{94,95}. Nevertheless, the aim of the present study is not reproducing exactly the AE stability of the discharge, a linear analysis of this case was done in previous communications \cite{73}, neither identifying the exact EP $\beta$ required to induce the transition from the soft to hard MHD limit. The goal here is rather an initial analysis of the mechanisms involved with a varying NBI heating input using a reduced selection of modes. In addition, including $n=1$ to $6$ modes in the simulations lead to the destabilization of multiple AEs that may show a partial overlapping during their evolution in the saturated phase, even for cases with a rather low EP $\beta$ value, thus the analysis of the system stability with respect to the soft and hard MHD limits cannot be readily performed. In the following, nonlinear simulations with different EP $\beta$, energy and radial density profiles are analyzed.

\section{Effect of the NBI operational regime on the AE saturation phase \label{sec:nonlinear}}

This section shows the outcome of a set of nonlinear simulations performed to analyze the effect of the NBI power injection, voltage and deposition region on the AE saturation phase. Optimization trends of the plasma heating efficiency are identified with respect to the expected EP transport induced in each configuration.

\subsection{NBI power injection}

The increment of the NBI power injection leads to a larger population of EPs in the plasma, thus the EP drive to trigger AEs is stronger. The simulations reproduce an enhancement of the NBI power injection as an increment of the EP $\beta$, that is to say, by an up scaling of the EP density at fixed EP energy. In the following, the simulation results for cases with an EP $\beta$ of $0.005$, $0.01$ and $0.02$ are discussed.

\begin{figure}[h!]
\centering
\includegraphics[width=0.5\textwidth]{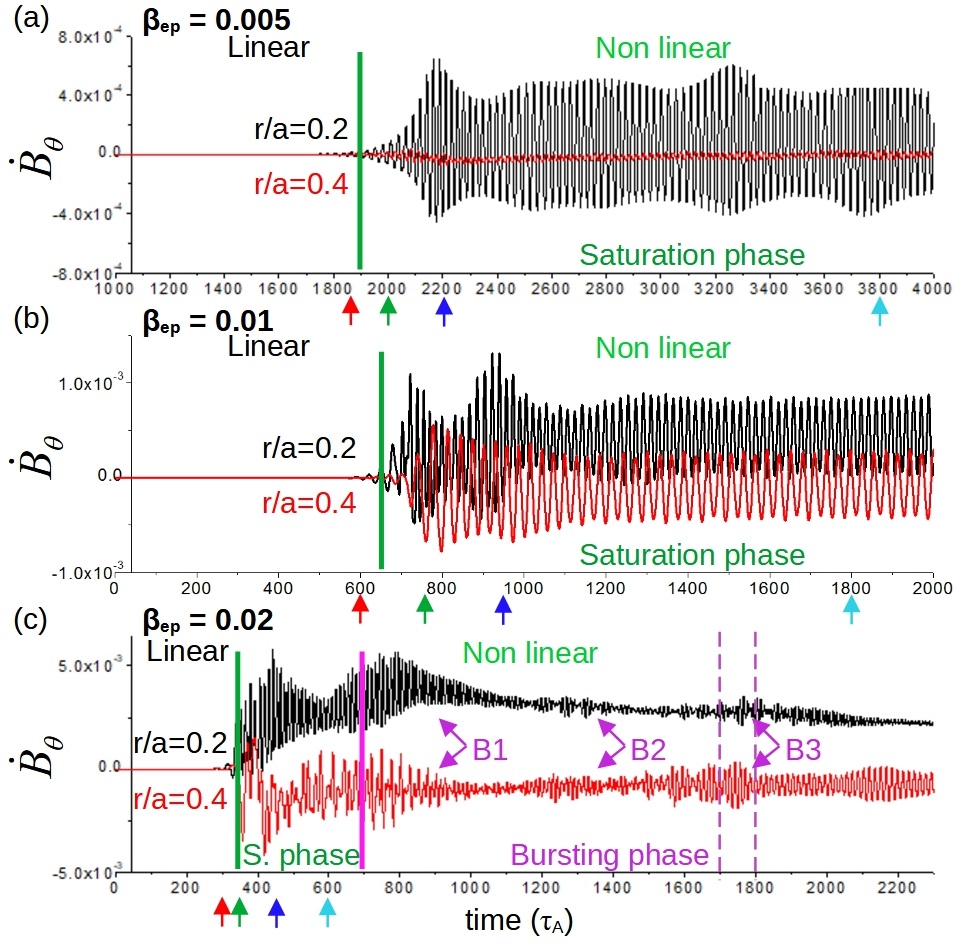}
\caption{Perturbation of the poloidal component of the magnetic field in the nonlinear simulations with an EP $\beta$ of (a) $0.005$, (b) $0.01$ and (c) $0.02$. The black line indicates the perturbation at $r/a =0.2$ and the red line the perturbation at $0.4$. The vertical bold green lines indicates the transition from the linear to the saturation phases. The vertical bold pink line indicates the beginning of the bursting phase (the bursts are indexed as B $+$ number). The vertical dashed pink line shows the bursting event analyzed in figure $6$. The colored arrows indicate the simulation time of the eigenfunctions plotted in figure $4$.}\label{FIG:3}
\end{figure}

\begin{figure}[h!]
\centering
\includegraphics[width=0.5\textwidth]{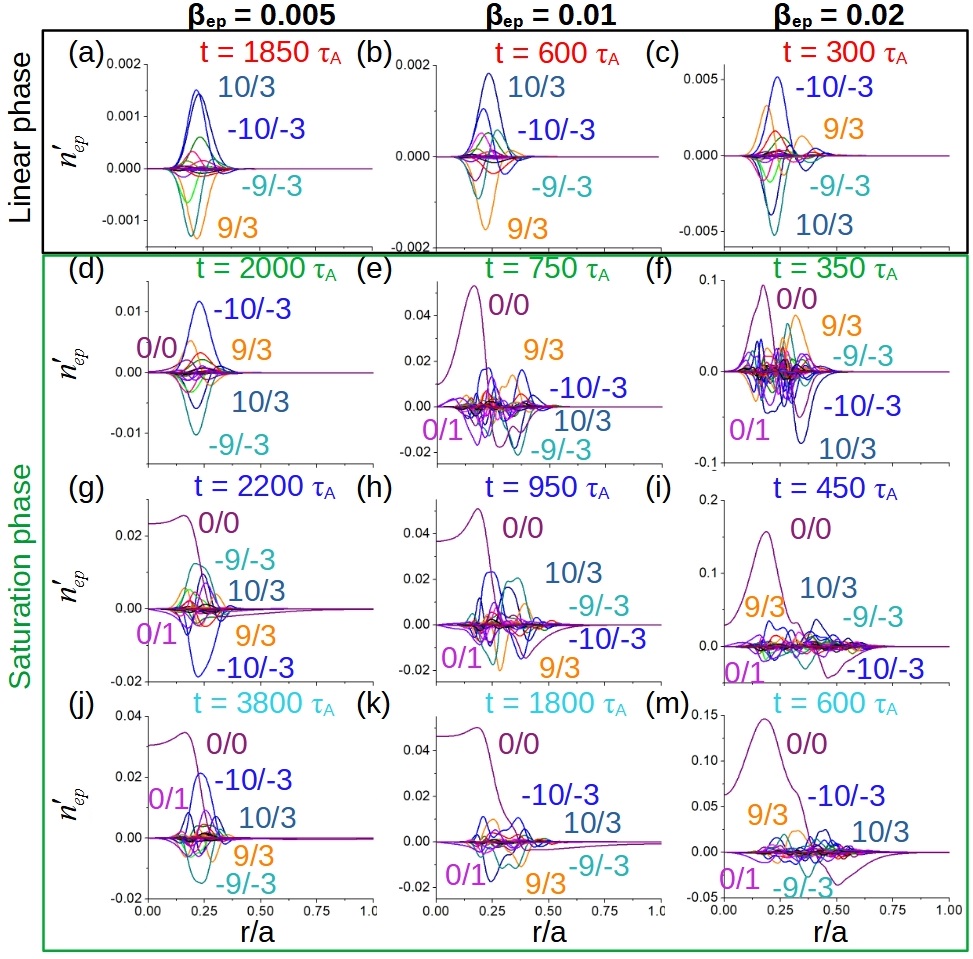}
\caption{Eigenfunctions of the EP density perturbation in the nonlinear simulations. Case with EP $\beta = 0.005$ during (a) the linear phase at $t=1850 \tau_{A0}$ and the saturation phase at (d) $t=2000 \tau_{A0}$, (g) $t=2200 \tau_{A0}$ and (j) $t=3800 \tau_{A0}$. Case with EP $\beta = 0.01$ during (b) the linear phase at $t=600 \tau_{A0}$ and the saturation phase at (e) $t=750 \tau_{A0}$, (h) $t=950 \tau_{A0}$ and (k) $t=1800 \tau_{A0}$. Case with EP $\beta = 0.02$ during (c) the linear phase at $t=300 \tau_{A0}$ and the saturation phase at (f) $t=350 \tau_{A0}$, (i) $t=450 \tau_{A0}$ and (m) $t=600 \tau_{A0}$.}\label{FIG:4}
\end{figure}

\begin{figure}[h!]
\centering
\includegraphics[width=0.5\textwidth]{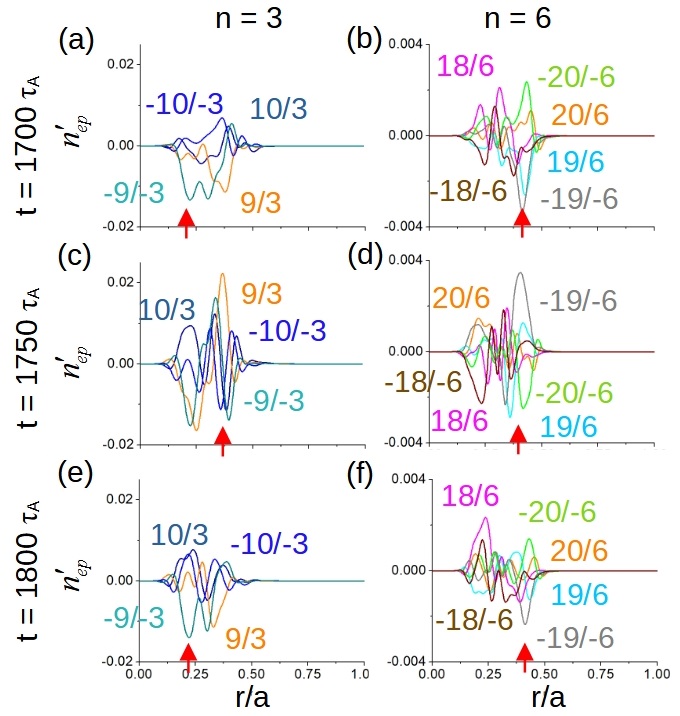}
\caption{Eigenfunctions of the EP density perturbation during the burst $B3$ in the simulation with EP $\beta = 0.02$. $n=3$ perturbation at (a) $t=1700 \tau_{A0}$, (c) $t=1750 \tau_{A0}$ and (e) $t=1800 \tau_{A0}$. $n=6$ perturbation at (b) $t=1700 \tau_{A0}$, (d) $t=1750 \tau_{A0}$ and (f) $t=1800 \tau_{A0}$. Red arrows indicate the radial location of the eigenfunction amplitude maxima.}\label{FIG:5}
\end{figure}

Figure~\ref{FIG:3} shows the evolution of the poloidal component of the magnetic field perturbation among the nonlinear simulations. The amplitude of the perturbations increases with the simulation EP $\beta$, as a consequence of a stronger EP drive. The simulation with EP $\beta = 0.005$ shows a perturbation localized near the magnetic axis (black line) whose amplitude remains almost constant in the saturation phase. The enhancement of the EP drive in the simulation with EP $\beta = 0.01$ leads to an AE that covers all the inner plasma region; this is the reason why the plasma at $r/a=0.4$ is also destabilized (red line). Nevertheless, the perturbation amplitude remains rather constant during the AE saturation phase. Large differences are observed in the simulation with EP $\beta = 0.02$ compared to the previous cases, particularly the emergence of a bursting phase characterized by the non steady evolution of the system.  The burst events are identified as a chain of perturbations correlated in time and amplitude at different plasma radial locations (indexed as B $+$ number). The bursts do not consist of a single system relaxation, there is a chain of relaxations that increase in amplitude until reaching a maximum and begin to decrease. The bursts identified in the figure~\ref{FIG:3} correspond to the largest amplitude perturbation of the poloidal component of the magnetic field, showing also the eigenfunction with the largest amplitude (fig.~\ref{FIG:4}), the largest overlapping between $n=3$ and $6$ modes (fig.~\ref{FIG:5}) as well as a local maximum of the growth rate calculated in linear simulations performed using the evolved profiles from the nonlinear simulations (fig.~\ref{FIG:13}).

Figure~\ref{FIG:4} indicates the eigenfunction of the EP density perturbation along the nonlinear simulations in the linear and saturation phases of the simulations. The eigenfunction during the linear phase shows a dominant $9/3-10/3$ TAE triggered in the inner plasma region whose amplitude grows as the EP $\beta$ of the simulation increases (panels a, b and c). Once the simulation enters in the saturation phase, the energy transfer from the TAE toward the thermal plasma causes the destabilization of the $0/0$ mode (panels d, e f). The amplitude of the $0/0$ mode is also larger as the EP $\beta$ of the simulation increases, pointing out that the perturbation of the thermal plasma is stronger as the TAE is further destabilized, that is to say, the energy transfer towards the thermal plasma is larger. The TAE does not stabilize during the saturation phase as a consequence of the enhanced EP transport and the partial loss of the EP population because the EP drive is set constant fixing the EP $\beta$ of the simulation. Consequently, the TAE amplitude is balanced between the EP drive and the energy transfer towards the thermal plasma leading to the generation of shear flows (panels g to i). It should be noted that the amplitude of the $0/0$ mode increases until the system reaches a cyclic evolution showing predator-prey features (panels j to m). This dynamic balance remains quasi-steady in the simulations with EP $\beta \leq 0.01$, although it is unable to last in the simulation with EP $\beta = 0.02$ and the system transits to the bursting phase.

Figure~\ref{FIG:5} shows the eigenfunction of the $n=3$ and $6$ EP density perturbation during the burst $B3$ triggered in the nonlinear simulations with EP $\beta = 0.02$ between $t=1700$ and $1800\tau_{A0}$ (see fig.~\ref{FIG:3} panel c, horizontal dashed pink lines). The strongest perturbation of the poloidal component of the magnetic field is observed at $t=1750 \tau_{A0}$ once the amplitude of the $9/3-10/3$ TAE and $18/6-20/6$ EAE are maxima (panel c and d). In addition, the local maxima of the $9/3-10/3$ TAE and $18/6-20/6$ EAE amplitude are located at the same radial location, in the middle plasma region ($r/a = 0.4$). On the other hand, the TAE local maxima are observed closer to the magnetic axis ($r/a = 0.2$) at $t=1700$ and $ 1800\tau_{A0}$ (panels a and e) although the EAE local maxima remain at $r/a = 0.4$ (panels b and f). Consequently, the triggering of the burst is caused by the radial overlapping of the $n=3$ TAE and $n=6$ EAE.

\begin{figure}[h!]
\centering
\includegraphics[width=0.5\textwidth]{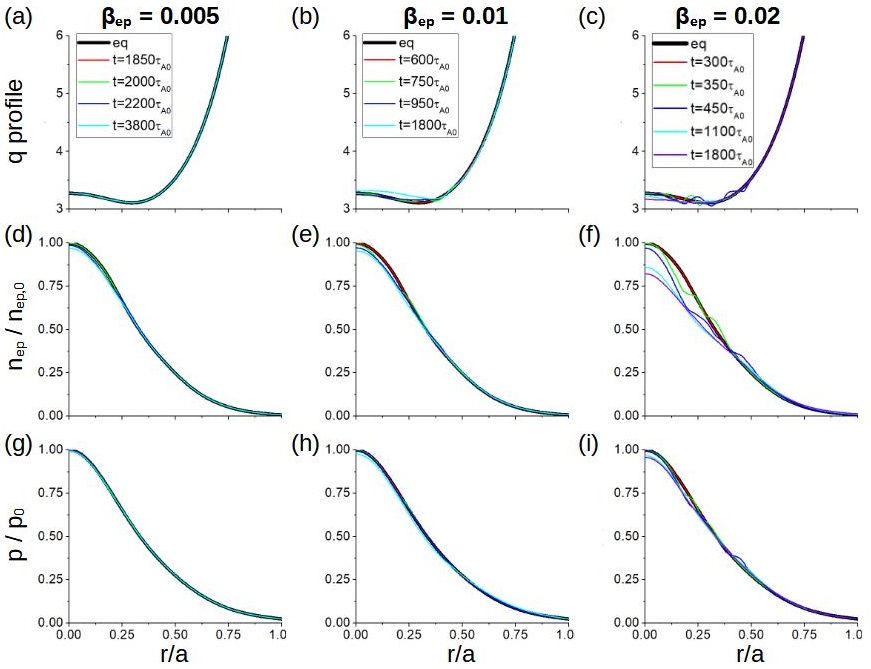}
\caption{Evolution of the equilibrium profiles. Simulation with EP $\beta = 0.005$: (a) safety factor, (d) EP density and (g) pressure profiles. Simulation with EP $\beta = 0.01$: (b) safety factor, (e) EP density and (h) pressure profiles. Simulation with EP $\beta = 0.02$: (c) safety factor, (f) EP density and (i) pressure profiles.}\label{FIG:6}
\end{figure}

Figure~\ref{FIG:6} indicates the evolution of the safety factor, EP density and pressure profiles along the nonlinear simulations for different EP $\beta$ values. The simulation for an EP $\beta = 0.005$ shows a negligible variation of the safety factor and pressure profiles (panels a and g) and a decrease of the EP density near the magnetic axis around $4 \%$ (panel d), pointing out the perturbation induced by the $9/3-10/3$ TAE on the thermal plasma is very small and the EP transport is rather small. The variation of the EP density profile is very localized thus the system relaxation corresponds to a system in the soft MHD limit. The profile evolution in the simulation with EP $\beta = 0.01$ indicate a stronger perturbation of the thermal plasma because the nonlinear energy transfer from the TAE is also enhanced, leading to the generation of more intense shear flows and zonal currents that modify the pressure and safety factor profiles, respectively (panels b and h). In addition, a larger decrease of the EP density is observed near the magnetic axis, around $6 \%$, thus the EP transport induced also increases. Even though the TAE perturbation enhances compared to the EP $\beta = 0.005$ case, the system still shows a relaxation that corresponds to the soft MHD limit. On the other hand, the evolution of the profiles in the simulation with EP $\beta = 0.02$ show important differences with the previous cases. The enforced destabilizing effect of the EP causes the destabilization of a strongly unstable TAE as well as the destabilization of the $n=6$ EAE, inducing even more intense shear flows and zonal currents that cause larger excursions of the pressure and safety factor profiles further away from the equilibrium values (panels c and i). Now, several regions of flattening in the safety factor and pressure profiles are observed at different radial locations caused by the $n=3$ and $6$ AEs at the same simulation time. The largest variation is observed in the EP density profile dropping between the inner and middle plasma region, indicating the plasma relaxation global due to the combined effect of the $n=3$ and $6$ AEs; thus we classify the system  as being in the hard MHD limit. In addition, the EP density profile decrease around $20 \%$ at the magnetic axis shows an important enhancement of the EP transport and a decrease of the plasma heating efficiency by the NBI.

\begin{figure}[h!]
\centering
\includegraphics[width=0.5\textwidth]{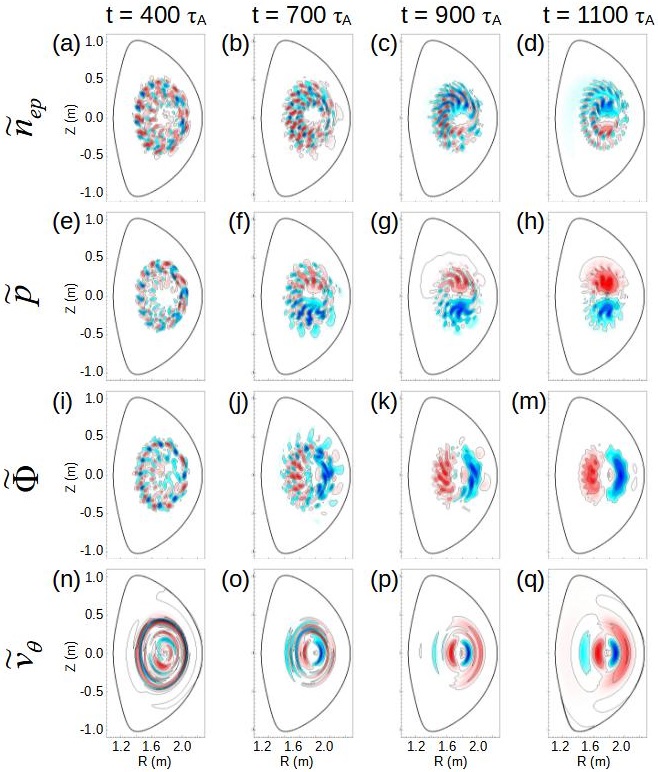}
\caption{Poloidal contours of the simulation with EP $\beta = 0.0$2 at $t=400$, $700$, $900$ and $1100\tau_{A0}$ between the magnetic axis and the middle plasma region ($r/a=0.6$). (a to d) EP density perturbation, (e to h) pressure perturbation, (i to m) electrostatic potential perturbation, (n to q) poloidal velocity perturbation.}\label{FIG:7}
\end{figure}

Figure~\ref{FIG:7} shows the contours at fixed poloidal angle of the EP density, pressure, electrostatic potential and thermal plasma poloidal velocity perturbations at different times along the simulation with EP $\beta = 0.02$. The poloidal contour of the EP density at the beginning of the saturation phase ($400\tau_{A0}$, panel a) indicates the perturbation induced by the $9/3-10/3$ TAE in the inner plasma region. The TAE perturbation along the bursting phase ($\geq 400\tau_{A0}$, panels b to d) show a weakening of the $10/3$ mode in the middle plasma region although an enhancement of the $9/3$ mode in the inner plasma region, that is to say, a transitional mode from a $9/3-10/3$ TAE and a $9/3$ RSAE. The poloidal contours of the pressure perturbation indicate the deformation of the flux surfaces induced by the AE. The flux surface deformation reaches a maximum at the beginning of the saturation phase (panel e), pointing out the rather large perturbation the AE induces in the thermal plasma. The flux surface deformation reduces in the bursting phase (panels e to h), thus the feedback between AEs and thermal plasma is also smaller and it is localized closer to the magnetic axis. The perturbation of the electrostatic potential is rather weak during the saturation phase (panel i) as well as the generation of zonal currents, mainly localized in the inner plasma region. On the other hand, during the bursting phase (panels j to m), the electrostatic potential perturbation is $5$ times stronger near the magnetic axis indicating the TAE-RSAE transitional mode leads to the generation of larger zonal currents, inducing the modifications in the safety factor profile (fig~\ref{FIG:6} panel c). The perturbation of the poloidal thermal velocity at the beginning of the saturation phase (panels n and q) indicates the TAE generates shear flows in the middle plasma region. The poloidal velocity perturbation increases during the bursting phase pointing out an enhancement of the shear flows between the magnetic axis and the middle plasma region. Such reinforcement of the shear flows in the bursting phase is linked to the system transition to a dynamic cyclic equilibrium with predator-prey features, that is to say, the intensity of the shear flows regulate the AE amplitude.

Summarizing, the NBI operational regime should avoid the transition from the soft to the hard MHD limit. If the NBI power injection is strong enough to operate in the hard MHD limit, the combined effect of multiple unstable AEs may induce global relaxations as bursting events, leading to a large enhancement of the EP transport and an important decrease of the NBI heating efficiency as well as potential damage on the plasma facing components. The simulation results are consistent with fast-ion loss detector (FILD) diagnostic measurements in DIII-D experiments that identified the generation of transient bursts above a given threshold of the NBI injection power, showing a higher recurrence as the beam power increases, also concluding that the bursts are caused by the AE overlapping leading to an enhanced EP diffusion and the triggering of avalanche-like events \cite{94}. Nevertheless, the simulations show that this issue can be partially avoided if the NBI is operated modifying the voltage to reduce the drive of the EP \cite{46}; this topic will be analyzed in the following section.

\subsection{NBI voltage}

This section is dedicated to an analysis of the effect of the EP energy variations, i. e. the NBI voltage, on the saturation phase of the AEs. Experiments and numerical analysis performed in DIII-D have already indicated a lower AE activity if the NBI voltage is reduced while keeping the power injection fixed \cite{46}. On the other hand, an enhancement of the AE activity is observed if the NBI voltage increases at a fixed NBI power injection. These operation scenarios can be analyzed if the energy of the target EP population in the simulation is increased or decreased at a fixed EP $\beta$ value, that is to say, a lower EP energy is compensated by a higher EP density. In the following, four different cases are analyzed: a low EP energy case with $T_{f} = 15$ keV and EP $\beta = 0.01$ (LTS, low temperature soft MHD limit case), a high EP energy case with $T_{f} = 25$ keV and EP $\beta = 0.01$ (HTS, high temperature soft MHD limit case), a low EP energy case with $T_{f} = 15$ keV and EP $\beta = 0.03$ (LTH, low temperature hard MHD limit case) and a high EP energy case with $T_{f} = 25$ keV and EP $\beta = 0.03$ (HTH, high temperature hard MHD limit case). Reducing the EP energy from $21.4$ keV to $15$ keV implies the EP density increases by $41 \%$ although increasing the EP energy to $25$ keV leads to a EP density $14 \%$ lower (fixed the EP $\beta$). Thus, increasing (decreasing) the EP energy leads to a lower (higher) EP drive with respect to a smaller (larger) amount of EPs in the plasma. In addition to the above macroscopic density changes with variations in EP energy, there are also local changes in the EP distribution function at the $v_{A0}/3$ resonant sideband coupling velocity. Increasing the energy from $21.4$ keV to $25$ keV decreases the distribution function by $8.3 \%$ at $v = v_{A0}/3$, while decreasing the energy to $15$ keV increases it by $14.2 \%$.

\begin{figure}[h!]
\centering
\includegraphics[width=0.5\textwidth]{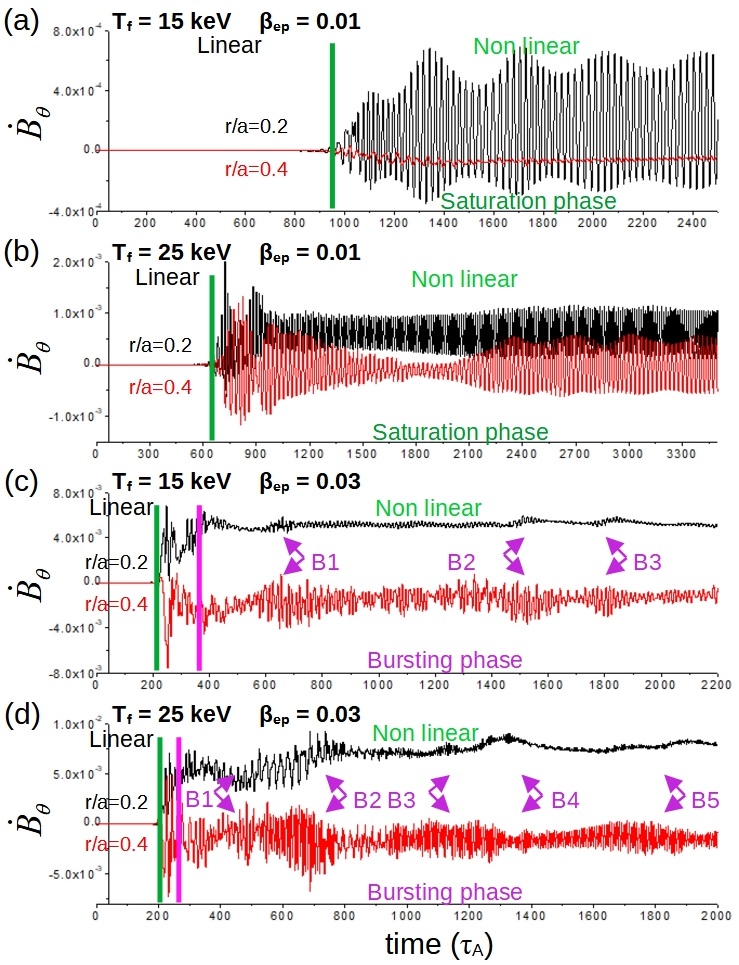}
\caption{Perturbation of the poloidal component of the magnetic field in the nonlinear simulations (a) LTS, (b) HTS, (c) LTH and (d) HTH. The black line indicates the perturbation at $r/a =0.2$ and the red line the perturbation at $0.4$. The vertical bold green line indicates the transit between the linear and nonlinear phases of the simulation. The vertical bold pink line indicates the beginning of the bursting phase (the bursts are indexed as B $+$ number).}\label{FIG:8}
\end{figure}

Figure~\ref{FIG:8} shows the evolution of the poloidal component of the magnetic field perturbation of simulations with different EP energy and $\beta$. If the panels (a) and (b) are compared to the panel (b) of figure~\ref{FIG:3}, the amplitude of the perturbation is smaller in the LTS case (almost $2$ times smaller) although larger in the HTS case (around $50 \%$ higher). In addition, the perturbation in the LTS case is localized nearby the magnetic axis, similar to the simulation with EP $\beta = 0.005$ (panel a of figure~\ref{FIG:3}) indicating that the width of the AE is smaller. That means a lower EP drive in the LTS case leads to important differences in the AE saturation phase compared to the HTS case even though both simulations show a plasma in the soft MHD limit. Same discussion can be done comparing LTH and HTH simulations, panels (c) and (d). The perturbation induced in the LTH case is around $30 \%$ smaller with respect to the HTH case. In addition, the number of bursting events induced in the HTH case is larger as well as perturbation amplitude at $r/a=0.2$ and $0.4$. That means reducing the voltage of an NBI operating in the hard MHD limit reduces the intensity and recurrence of the burst events because the EP drive is lower.

\begin{figure}[h!]
\centering
\includegraphics[width=0.5\textwidth]{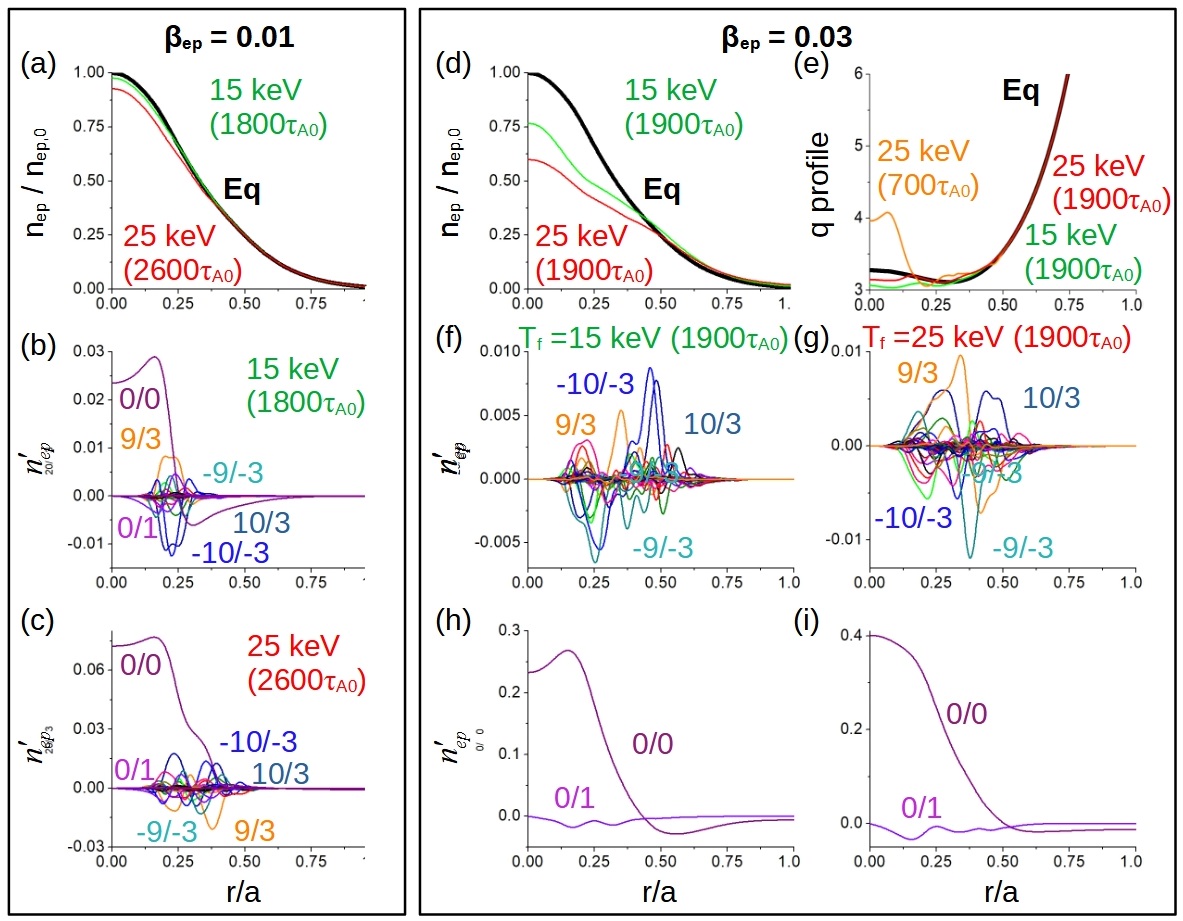}
\caption{(a) EP density profiles in the simulations LTS (green) and HTS (red) at the late saturation phase. Eigenfunction of the EP density perturbation at the late saturation phase of the (b) LTS and (c) HTS simulations. (d) EP density profiles in the simulations LTH (green) and HTH (red) at the late saturation phase. (e) Safety factor profiles in the simulations LTH in the late bursting phase (green),  and HTH at the early (orange) and late (red) bursting phase. Eigenfunction of the EP density perturbation (without $n=0$ components) at the late bursting phase of the (f) LTH and (g) HTH simulations. Eigenfunction of the EP density perturbation (only $n=0$ components) at the late bursting phase of the (h) LTH and (i) HTH simulations.}\label{FIG:9}
\end{figure}

Figure~\ref{FIG:9} indicates the evolution of the EP density and safety factor profiles as well as the eigenfunctions of the EP density perturbation in the simulations LTS, HTS, LTH and HTH. The panel (a) shows a decrease of the EP density profile of $3 \%$, half compared to the reference simulation with EP $\beta = 0.01$ (figure ~\ref{FIG:6}, panel (e)) and slightly smaller with respect to the simulation with EP $\beta = 0.005$ (figure ~\ref{FIG:6}, panel (e)). That means a reduction of the NBI voltage not only reduces the EP drive, but also the EP transport induced by a NBI operating at a lower voltage is smaller compared with a NBI operating with a weaker power injection. The decrease of the EP drive in the LTS simulation compared to the HTS case is demonstrated in the lower amplitude of the $9/3-10/3$ TAE and $0/0$ mode, panels b and c, pointing out the TAE is less unstable and the energy transfer towards the thermal plasma is lower. In addition, the amplitude of the $9/3-10/3$ TAE and $0/0$ mode in the LTS simulations is smaller compared to the reference simulations with EP $\beta = 0.01$ and $0.005$. The EP density profile drops $8 \%$ at the magnetic axis in the HTS simulation, showing a more radially extended profile decrease compared to the LTS case, indicating the destabilization of a wider TAE in the inner plasma region as can be observed in the panel b of figure~\ref{FIG:8}. The large decrease of the EP density profile between the magnetic axis and the middle plasma region in the simulations LTH and HTH indicates the NBI operates in the hard MHD limit (panel d). The EP density drop at the magnetic axis is larger in the HTH simulation, $40 \%$, compared to the LTH case, $23 \%$. It should be noted that the profile decrease in the reference simulation with EP $\beta = 0.02$ is around $20 \%$, thus the reduction of the EP transport observed by decreasing the NBI voltage is similar compared with an NBI operation with lower power injection. Consequently, reducing the NBI voltage is less efficient to improve the plasma heating efficiency in the hard MHD limit with respect to the soft MHD limit. The evolution of the safety factor profile, panel e, shows large excursions from the equilibrium values in the HTH case (orange and red lines), pointing out the generation of rather intense zonal currents, particularly nearby the magnetic axis. On the other hand, the evolution of the safety factor profile in the LTH simulation is smoother. In both simulations the effect of the zonal currents tend to reduce the reverse shear region by decreasing the safety factor profile nearby the magnetic axis. Panels f and g demonstrate the stronger EP drive in the simulation HTH compared to the LTH case, leading to a TAE amplitude $30 \%$ larger. The same discussion can be done with respect to the amplitude of the $0/0$ mode, panels h and i, $32 \%$ larger in the HTH case indicating a strong destabilization of the thermal plasma with respect to the LTH simulation.

In summary, reducing the NBI voltage leads to a decrease of the EP drive that affects the AE saturation phase. The AE perturbation is more radially located nearby the magnetic axis and the EP transport induced is lower. In addition, if the NBI operates in the soft MHD limit, there is a larger reduction of the EP transport if the NBI voltage is reduced compared with an operational regime decreasing the power injection. Consequently, an improvement of the plasma heating by reducing the NBI voltage is more efficient compared to a reduction of the NBI power injection. On the other hand, such optimization trend is not as robust in the hard MHD limit as in the soft MHD limit.

\subsection{NBI radial deposition region}

\begin{figure}[h!]
\centering
\includegraphics[width=0.3\textwidth]{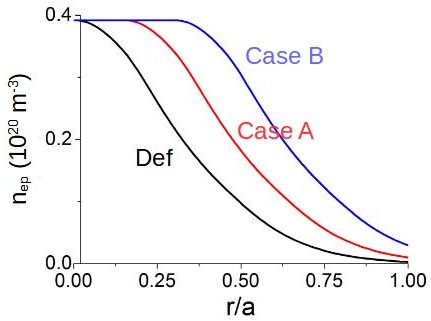}
\caption{EP density profiles in the reference case (black line), case A (red line) and case B (blue line).}\label{FIG:10}
\end{figure}

This section is dedicated to study the effect of the NBI radial deposition region on the AE saturation phase. There is experimental evidence in DIII-D plasma showing a reduction of the AE activity in the inner plasma region during off-axis NBI operations \cite{45,96}. An off-axis NBI injection can be reproduced in the first approximation as an outwards radial displacement of the EP density profile, fixing the gradient of the profile for simplicity. A set of nonlinear simulations are performed dedicated to analyze the AE saturation phase if the NBI is deposited on-axis (reference case) and the NBI is deposited further off-axis, particularly at $r/a=0.15$ (case A) and $0.3$ (case B). Figure~\ref{FIG:10} shows the EP density profiles of each simulation. The EP $\beta$ of the simulations is $0.03$, leading to a robust NBI operational regime in the hard MHD limit in the reference case (consistent with the analysis performed with respect to the EP $\beta$ showing a transition from the soft to the hard MHD limit if the EP $\beta = 0.02$).

\begin{figure}[h!]
\centering
\includegraphics[width=0.5\textwidth]{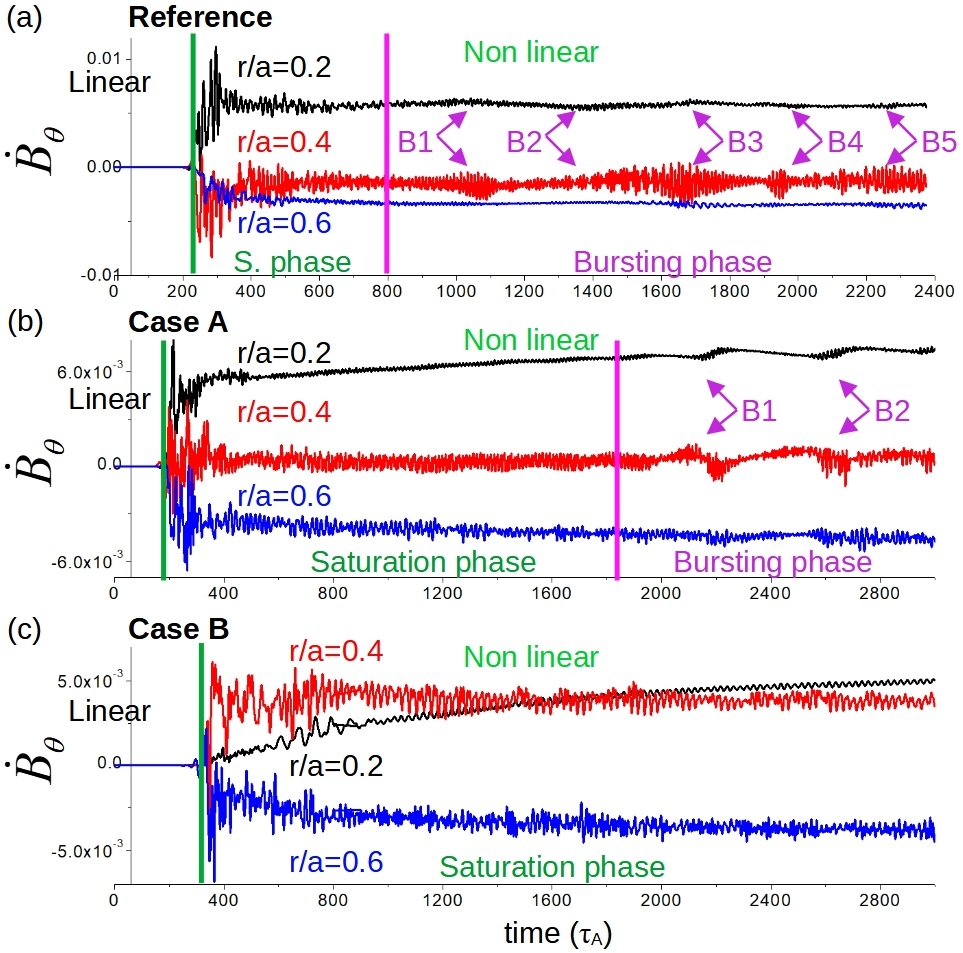}
\caption{Perturbation of the poloidal component of the magnetic field in the nonlinear simulations with (a) on-axis NBI injection (reference case), (b) off-axis NBI injection at $r/a=0.15$ (case A) and (c) off-axis NBI injection at $r/a=0.3$ (case B). The black line indicates the perturbation at $r/a =0.2$, the red line at $0.4$ and the blue line at $0.6$. The vertical bold green line indicates the transit between the linear and nonlinear phases of the simulation. The vertical bold pink line indicates the beginning of the bursting phase (the bursts are indexed as B $+$ number).}\label{FIG:11}
\end{figure}

Figure~\ref{FIG:11} shows the evolution of the poloidal component of the magnetic field perturbation in simulations with different NBI radial deposition region. The largest amplitude perturbation is observed in the on-axis case, $30 \%$ larger compared to the off-axis case A and $40 \%$ larger with respect the off-axis case B. Consequently, the EP drive decreases as the NBI is located further away from the magnetic axis. In addition, the length of the saturation phase in the off-axis case A is $2$ times longer compared to the reference case. The recurrence and the burst perturbation amplitude is also larger in the reference case. On the other hand, the perturbation in the reference case is located between the magnetic axis and inner plasma region ($r/a = 0.4$) although the perturbation in the off-axis case A is wider, covering the plasma from the magnetic axis to the middle plasma region ($r/a=0.6$). It should be noted that the transition to the bursting phase is not observed in the off-axis case B, pointing out the EP drive is not large enough to induce the transition to the hard MHD limit. Also, the perturbation is observed in the middle plasma region, thus the AE is radially localized indicating that the NBI operation is in the soft MHD limit.

\begin{figure}[h!]
\centering
\includegraphics[width=0.5\textwidth]{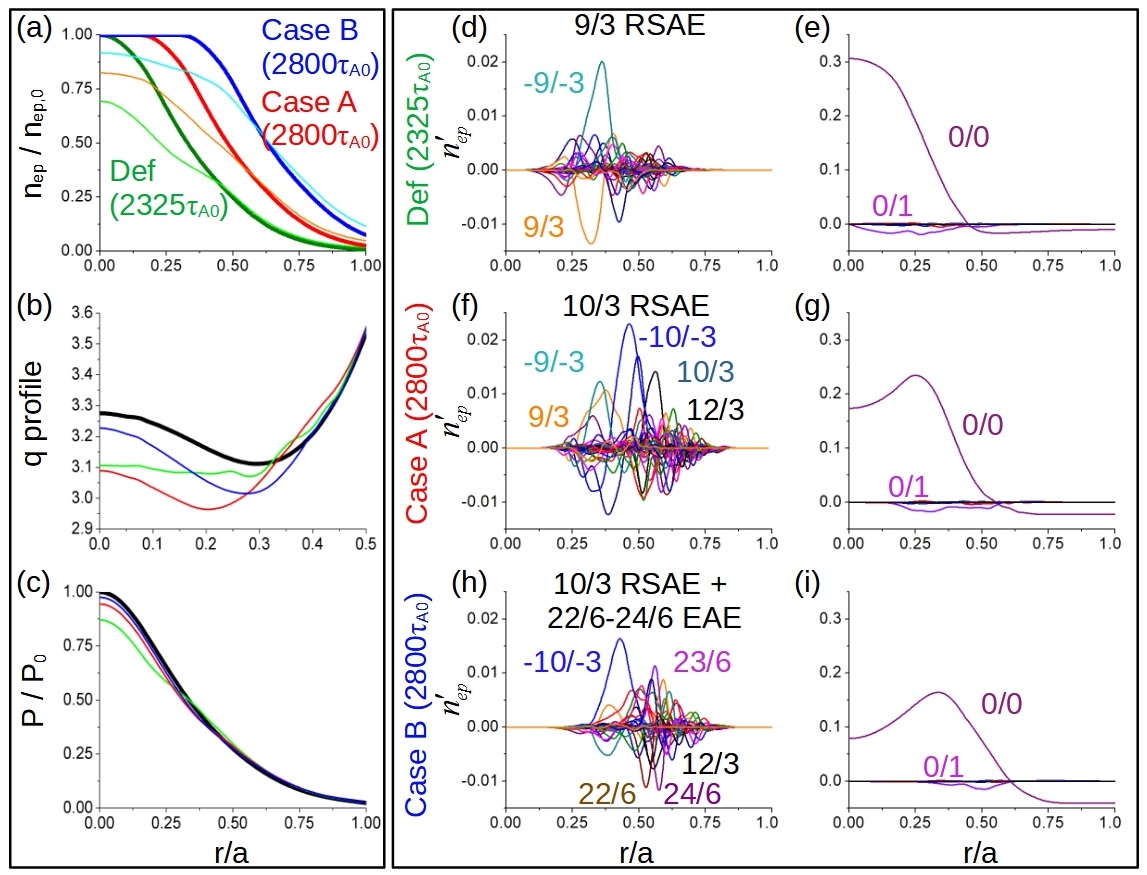}
\caption{Evolution of the (a) EP density, (b) safety factor and (c) pressure profiles in simulations with different NBI radial deposition regions. The bold green, red and blue lines indicate the equilibrium EP density profile in the reference, case A and case B, respectively. The light green, orange and cyan lines indicate the evolved EP density profile in the reference, case A and case B, respectively. The bold black lines show the equilibrium safety factor and pressure profiles. Eigenfunction of the EP density perturbation in the on-axis simulation (d) without $n=0$ components and (e) only $n=0$ components. Eigenfunction of the EP density perturbation in the off-axis case A simulation (f) without $n=0$ components and (g) only $n=0$ components. Eigenfunction of the EP density perturbation in the off-axis case B simulation (h) without $n=0$ components and (i) only $n=0$ components.}\label{FIG:12}
\end{figure}

Figure~\ref{FIG:12} shows the evolution of the EP density, safety factor and pressure profiles as well as the eigenfunctions of the EP density perturbation in simulations for a NBI deposited at different radial locations. The EP density profile, panel (a), indicates the largest decrease at the magnetic axis for the reference case, around $30 \%$, as well as a drop of $18 \%$ in case A and $7 \%$ in case B. The total amount of EP losses in the reference case, calculated comparing the areas below equilibrium and evolved profiles, is $21 \%$, although $15 \%$ for case A and $7 \%$ for case B, indicating a relative larger EP transport in the reference model. On the other hand, it must be considered that the area enclosed by the EP density profile is different in each simulation, implying a different integrated EP density for each case. The number of EPs is $1.4$ times larger in the case A and $1.79$ in case B with respect to the reference case, thus the absolute amount of EP lost in the reference and case A is very similar, $21 \%$, although it is lower in the case B, approximately $13 \%$. Consequently, even though the EP drive in the off-axis case A is lower compared to the reference case, the AE activity induces similar EP transport because the AE is more radially extended, thus an off-axis NBI operation may not lead to an improved heating efficiency if the beam is deposited at $r/a \leq 0.15$. The evolution of the safety factor profile, panel b, indicates on-axis NBI operation tends to reduce the reserve shear region in the inner plasma region by decreasing the safety factor near the magnetic axis due to the effect of the zonal currents induced by the AEs. On the other hand, off-axis NBI operation tends to increase the depth of the reverse shear region as the NBI is deposited further outwards. The perturbation induced in the pressure profile is rather small in the off axis cases, panel c, although the pressure profile at the magnetic axis decreases $13 \%$ in the on-axis case, pointing out a stronger perturbation of the thermal plasma in the reference case. The dominant mode in the on-axis simulation is a $9/3$ RSAE triggered at the inner plasma region ($r/a = 0.3$), panel d, although the dominant mode in the off-axis case A is a$10/3$ RSAE unstable at the middle plasma, panel f, and a combined $10/3$ RSAE $+$ $22/6-24/6$ EAE at the middle plasma region in the off-axis case B, panel h. The $9/3$ RSAE triggered in the on-axis case shows the largest amplitude followed by the $10/3$ RSAE in the off-axis case A. The $10/3$ RSAE $+$ $22/6-24/6$ EAE show the smallest amplitude in the off-axis case B. The amplitude an radial location of the $0/0$ and $0/1$ modes shows the on-axis case cause to the strongest perturbation of the thermal plasma in the inner plasma, decreasing and moving outwards as the NBI is deposited further off-axis (panels e, g and i). 

In conclusion, performing an off-axis NBI injection reduces the drive of the EP in the plasma and the amplitude of the perturbation, although the fraction of the plasma destabilized extends from the inner plasma towards the middle plasma region. The analysis indicates the EP transport in on-axis and off-axis cases is rather similar if the NBI is deposited at $r/a \leq 0.15$, although around $40 \%$ smaller if the NBI is deposited at $r/a = 0.3$. Also, the on-axis NBI injection tends to reduce the reverse shear region in the inner plasma region due to the generation of zonal currents by the RSAE. On the other hand, off-axis NBI operation may cause the opposite effect, inducing a deepening of the reverse shear region. That means, off-axis NBI operations may lead to an improved plasma heating if the beam is deposited further away from $r/a = 0.15$, leading to a configuration with reduced EP driven and transport. The simulations reproduce similar trends compared to DIII-D experiments that measured a weakening of RSAE and core TAEs if the NBI is injected off-axis \cite{96}.

\section{Effect of the equilibrium profiles evolution on the AE linear stability \label{sec:linear2}}

This section is dedicated to analyze the effect of the equilibrium profile evolution obtained along the nonlinear simulations on the linear AE stability. The analysis method consist on performing a set of linear stability evaluations, based on the perturbed profiles at different times along the nonlinear simulation for different EP $\beta$ values (cases with EP $\beta = 0.005$, $0.01$, $0.02$ and $0.03$).

\begin{figure}[h!]
\centering
\includegraphics[width=0.5\textwidth]{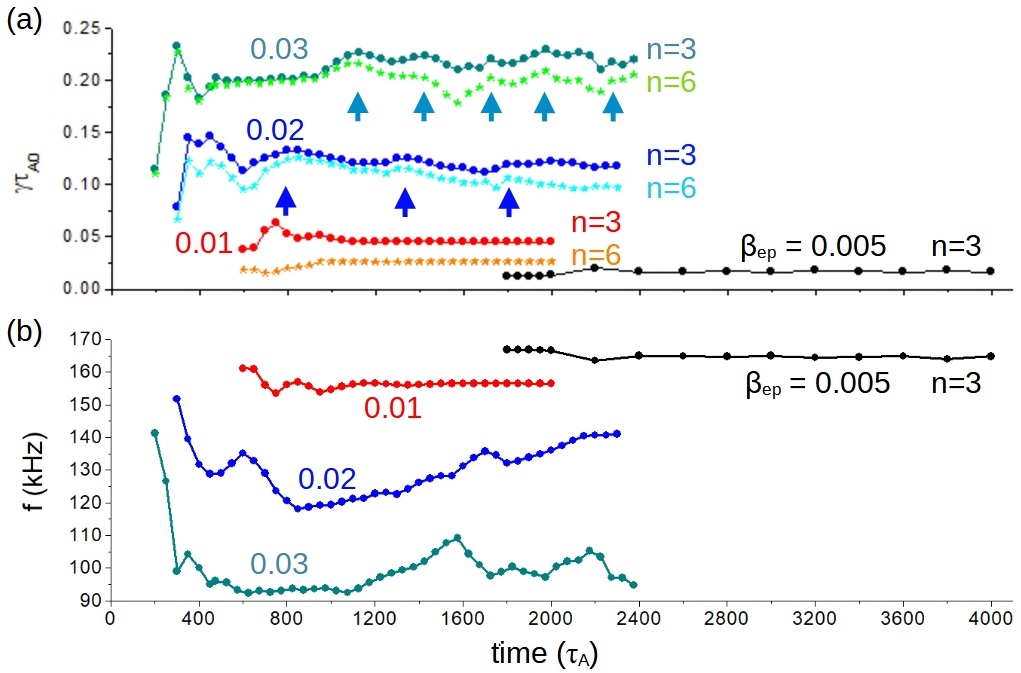}
\caption{Linear AE stability analysis using the evolved profiles of the nonlinear simulations. (a) AE growth rate and (b) frequency. The dots indicate the nonlinear simulation times included in the study (using the corresponding evolved equilibrium profiles). The black lines represents the $n=3$ AE in the EP $\beta=0.005$ case, the red (orange) line the $n=3$ ($n=6$) AE in the EP $\beta=0.01$ case, the blue (cyan) line the $n=3$ ($n=6$) AE in the EP $\beta=0.02$ case and the green (light green) line the $n=3$ ($n=6$) AE in the EP $\beta=0.03$ case. The blue (cyan) bold arrows indicate the destabilization time of the burst events in the EP $\beta = 0.02$ ($0.03$) simulations.}\label{FIG:13}
\end{figure}

Figure~\ref{FIG:13} indicates the linear growth rate and frequency of the dominant $n=3$ and $6$ AEs using the evolved equilibrium profiles. The time averaged growth rate using the evolved profiles of EP $\beta = 0.03$ simulation is around $2$ times larger compared to EP $\beta = 0.02$ case, $4$ times larger with respect to the EP $\beta=0.01$ case and $8$ times bigger regarding EP $\beta=0.005$ simulation, consistent with the linear dependency of the EP $\beta$ with $n=3$ and $6$ AE growth rates (panel (a)). Nevertheless, the AE growth rates show important oscillations during the simulations; these are associated with the evolution of the equilibrium profiles induced by the AEs nonlinear feedback in the thermal plasma. There is a sharp increase of the AEs growth rate once the simulations enter in the saturation phase compared to the linear phase, stronger as the EP $\beta$ of the simulation increases. This is explained by the evolution of the EP density profile caused by the EP transport induced by the AE, leading to an reinforced EP density profile gradient in the inner plasma region. Such an effect can be observed in the panel f of the fig.~\ref{FIG:6} (simulation with EP $\beta = 0.02$) during the saturation phase, green and violet lines at $t=350$ and $450 \tau_{0A}$. In addition, if the AE is strong enough to cause a large perturbation of the EP density profile, the dominant mode observed in the linear and saturation phase could be different. This is the case of the simulation with EP $\beta = 0.03$, showing a dominant $9/3-10/3$ TAE during the linear phase although a dominant $9/3$ RSAE in the saturation and busting phases (see fig.~\ref{FIG:12} panel (d)). It should be noted that the $n=3$ AE is the fastest growing mode in all the simulations, although the growth rate of the $n=6$ AE is similar to the $n=3$ at the end of the saturation phase and beginning of the bursting phase. That means the bursting phase is triggered once the combined destabilizing effect of $n=3$ and $6$ AEs is maxima, inducing the largest perturbation on the thermal plasma. In addition, the EP $\beta = 0.01$ case shows the destabilization of the $n=6$ AE during the simulation saturation phase with a growth rate half of the $n=3$ AE, although the linear analysis using the equilibrium profiles found the $n=6$ AE is stable (see fig.~\ref{FIG:2} panel (b)). That means the $n=6$ AE is non-linearly destabilized by the combined effect of the nonlinear energy transfers from the $n=3$ AE towards the $n=6$ AE, the perturbation induced by the $n=3$ AE in the thermal plasma and the reinforced EP density profile gradient linked to the EP transport induced by the $n=3$ AE. It should be noted that the bursting events observed in the nonlinear simulations are correlated with a local maxima of the $n=3$ and $6$ AEs growth rate (blue and cyan arrows in the simulations with EP $\beta = 0.02$ and $0.03$, respectively), thus the burst events are caused by an enhanced overlapping of the AEs as the modes amplitude increases. The $n=3$ AE frequency (panel (b), $n=6$ AE frequency not shown) also changes due to the evolution of the equilibrium profiles, particularly if the dominant mode switches between linear and the saturation simulation phases. The most clear example is provided by the nonlinear simulation with EP $\beta = 0.03$, showing a decrease of the AE frequency from $145$ kHz in the linear phase to $95$ kHz during the saturation phase, followed by a succession of local maximum and minima linked to the destabilization bursting events.

\begin{figure}[h!]
\centering
\includegraphics[width=0.5\textwidth]{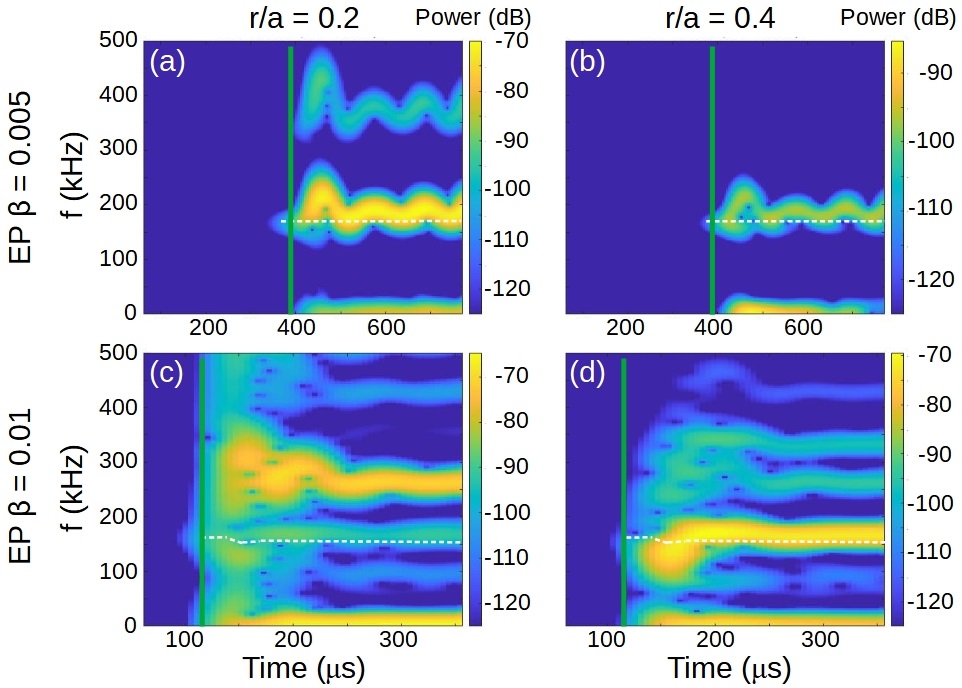}
\caption{Spectrogram of the poloidal magnetic field perturbation in the simulation with EP $\beta=0.005$ at (a) the magnetic axis ($r/a=0.2$) and (b) the inner plasma region ($r/a=0.4$). Same analysis for the simulation with EP $\beta=0.01$ at (c) the magnetic axis and (d) the inner plasma region. The dashed white line indicates the frequency range of the dominant AE calculated in the linear simulations using the evolved equilibrium profiles. The vertical bold green line indicates the AE transition from the linear to the saturated phase.}\label{FIG:14}
\end{figure}

Figure~\ref{FIG:14} shows the spectrograms of the poloidal magnetic field perturbation at $r/a = 0.2$ and $0.4$ for the simulations with EP $\beta=0.005$ and $0.01$, the cases in the soft MHD limit. This synthetic diagnostic is particularly useful because it allows comparing directly simulation results and the measured magnetic data in the experiments. The dashed white line over plots the AE frequency calculated in the linear simulations using the evolved equilibrium profiles of the nonlinear simulations. The strongest AE activity in the simulation with EP $\beta = 0.005$ is observed oscillating in the frequency ranges between $140$ and $230$ kHz nearby the magnetic axis (fig~\ref{FIG:14}, panels a and b). Overtones of the dominant AE activity are observed at frequency ranges above $300$ kHz (EAE and NAE gaps) and below $30$ kHz (BAE gap), although the EAE/NAE activity is only observed nearby the magnetic axis and the BAE activity cover all the inner plasma region. The analysis of the EP $\beta = 0.01$ case (fig~\ref{FIG:14}, panels (c) and (d)) shows the strongest AE activity is located at $r/a=0.4$ in the frequency range of $100$ to $190$ kHz. New overtones appears in the frequency range of $250$ kHz (EAEs particularly strong nearby the magnetic axis), NAE at $420$ kHz, $RSAE$ at $100$ kHz and $BAE$ below $30$ kHz. If the AE activity in both nonlinear simulations is compared to the results obtained by the linear simulations using the evolved profiles, the linear simulations calculate a similar frequency range for the dominant AE activity. Nevertheless, the AE frequency chirping induced by the nonlinear energy transfer towards the thermal plasma and $n=6$ modes cannot be reproduced by the linear simulations. Consequently, linear simulations using the evolved equilibrium profiles can provide a good approximation of the AE averaged activity during the saturation phase of the AEs if the plasma is in the soft MHD regime.

\begin{figure}[h!]
\centering
\includegraphics[width=0.5\textwidth]{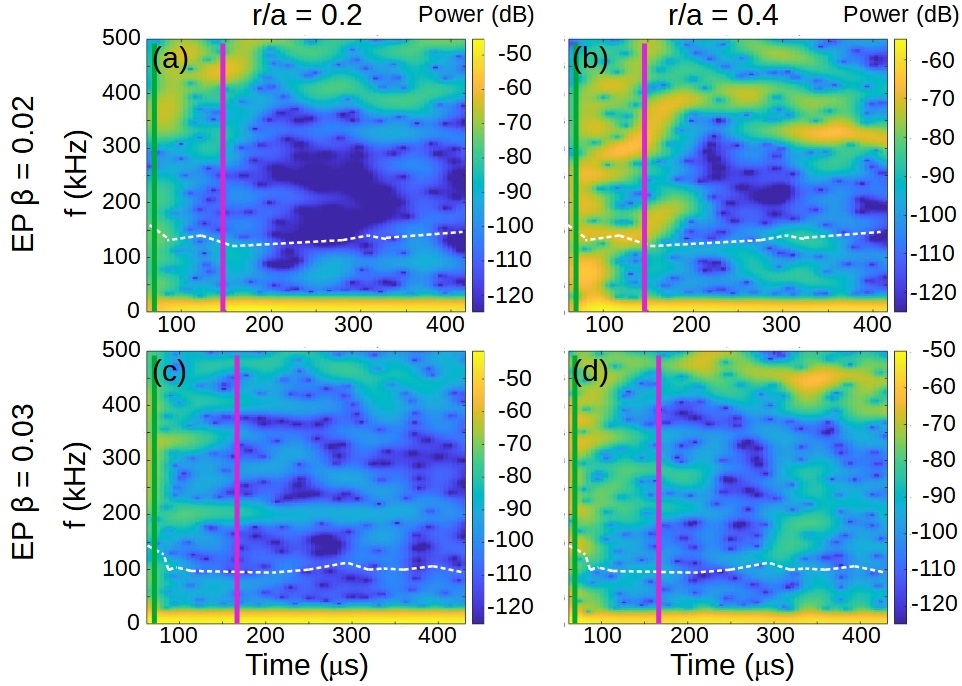}
\caption{Spectrogram of the poloidal magnetic field perturbation in the simulation with EP $\beta=0.02$ at (a) the magnetic axis ($r/a=0.2$) and (b) the inner plasma region ($r/a=0.4$). Same analysis for the simulation with EP $\beta=0.03$ at (c) the magnetic axis and (d) the inner plasma region. The dashed white line indicates the frequency range of the dominant AE calculated in the linear simulations using the evolved equilibrium profiles. The vertical bold green (pink) line indicates the AE transition from the linear to the saturation (bursting) phase.}\label{FIG:15}
\end{figure}

Figure~\ref{FIG:15} shows the spectrograms of the poloidal magnetic field perturbation at $r/a = 0.2$ and $0.4$ for the simulations with EP $\beta=0.02$ and $0.03$, the cases in the hard MHD limit. The spectrograms show a more complex interplay between $n=3$ and $6$ AEs as well as with the thermal plasma compared to the simulations in the soft MHD limit, leading to a large spreading of the AE activity in frequency and the destabilization of multiple overtones, particularly after the system enters in the bursting phase (bold pink line). It is unclear what the frequency range and radial location is of the dominant AE activity during the bursting phase, although the data tendencies may indicate a large activity of high frequency AEs above $300$ kHz at $r/a=0.4$ and low frequency modes at $r/a=0.2$. If the spectrograms of the EP $\beta = 0.02$ case, panels (a) and (b), are compared with the frequency of the dominant AE in the linear simulations using the evolved equilibrium profile (i. e., dashed white line), the linear simulation may reproduce the largest AE activity observed during the saturation phase around $120$ kHz at $r/a=0.4$. Nevertheless, the spectrogram also shows AE activity in the frequency range around $250$ to $300$ kHz as well as below $30$ kHz as strong as the activity at $120$ kHz. These components were not identified as the dominant AE activity in the linear simulations and there are likely nonlinearly driven. On the other hand, the spectrograms at the beginning of the bursting phase indicate an up-shift of the dominant AE activity not observed in the linear simulations, leading also to an inconsistent identification of the dominant AE activity. The same discussion applies to the simulation with EP $\beta = 0.03$, panels (c) and (d), showing an even large discrepancy with respect to the dominant AE activity during saturation and bursting phases between nonlinear simulations and linear simulations using evolved equilibrium profiles. That means the linear simulations using evolved equilibrium profiles cannot be used to reproduce the dominant AE activity of systems in the hard MHD limit. The nonlinear couplings of $n=3$ and $6$ AEs between them and the thermal plasma dominate the system evolution and must be included in the analysis to obtain a correct prediction of the AE activity.

In summary, the evolution of the equilibrium profiles and zonal flows / currents caused by the AE perturbation of the thermal plasma and EP density profile can lead to important differences in the AE stability comparing the linear and saturated phase of the nonlinear simulations. Thus, discrepancies between the prediction of linear and nonlinear simulations may appear, larger as the EP drive increases. For example, the time averaged growth rate of the $n=3$ AE calculated using the evolved equilibrium profiles of the nonlinear simulation with EP $\beta = 0.03$ is $0.22$, although the linear simulation using non evolved equilibrium profiles is $0.115$, almost two times smaller. On the other hand, the same analysis applied to the EP $\beta = 0.01$ nonlinear simulation provides a time averaged growth rate of $0.045$ although $0.03$ for the linear using non evolved equilibrium profiles. Thus, the linear simulations with non evolved profiles underestimate the AE growth rate in the AE saturation phase. A similar discussion applies for the AE frequency. The linear study performed in section \ref{sec:linear} compared to the linear analysis using the evolved equilibrium profiles indicate an over estimation of the AE frequency in the simulation without evolved equilibrium profiles, $7 \%$ lower if the EP $\beta = 0.005$, $11 \%$ in the EP $\beta = 0.01$, $23 \%$ in the EP $\beta = 0.02$ and $38 \%$ in the EP $\beta = 0.03$. Consequently, linear and nonlinear simulations results show large differences if the NBI operates in the hard MHD limit because the EP transport is enhanced by the effect of the resonances radial overlapping. If the NBI operates in the soft MHD regime, linear and nonlinear simulations show closer results, except for the nonlinear triggering of the stable mode in the linear analysis without evolved profiles, although such modes have a growth rate significantly smaller compared with the dominant modes, thus their impact in the plasma stability is rather limited. That means the linear analysis provides a reasonable first approximation of the AE stability if the EP drive is not strong enough to trigger global plasma relaxations.

\section{Conclusions and discussion \label{sec:conclusions}}

The effect of the NBI operational regime on the saturation phase of AEs is analyzed by performing a set of linear and nonlinear simulations using the gyro-fluid code FAR3d. The NBI operational regime is analyzed with respect to the power injection (EP $\beta$), voltage (energy of the target EP population) and radial deposition region (EP density profile). The AE stability in the linear and nonlinear simulations is compared calculating the linear AE growth rate and frequency using the evolved equilibrium profiles obtained in the nonlinear simulations.

The linear analysis using the non evolved equilibrium profiles indicates the fastest growing mode is triggered by the $n=3$ toroidal family, leading to the destabilization of a $9/3-11/3$ EAE with $f=160$ kHz (if the EP $\beta = 0.03$). The EP $\beta$ threshold for the destabilization of the $n=1$ to $4$ AEs is lower than $0.005$, although it should be at least $0.01$ to trigger an $n=5$ AE and $0.02$ to destabilize an $n=6$ AE. 

The nonlinear study of the NBI power injection on the AE stability indicates a transition from the soft to the hard MHD regime if the EP $\beta$ of the simulation is at least $0.02$, leading to global plasma relaxations identified as burst events caused by the radial overlapping of multiple EP resonances. The EP transport in the hard MHD limit largely increases compared to the simulations in the soft MHD regime, leading to a decrease of the EP density profile between the magnetic axis and the middle plasma region. Consequently, NBI operational regimes in the hard MHD limit should be avoided to improve the plasma heating performance and avoid damage on the plasma facing components of the device. It should be noted that these simulation results are consistent with DIII-D experimental observations that show an enhancement of the measured EP transport due to the overlapping of multiple AEs in discharges with variable NBI injection intensity \cite{94}.

\begin{figure}[h!]
\centering
\includegraphics[width=0.5\textwidth]{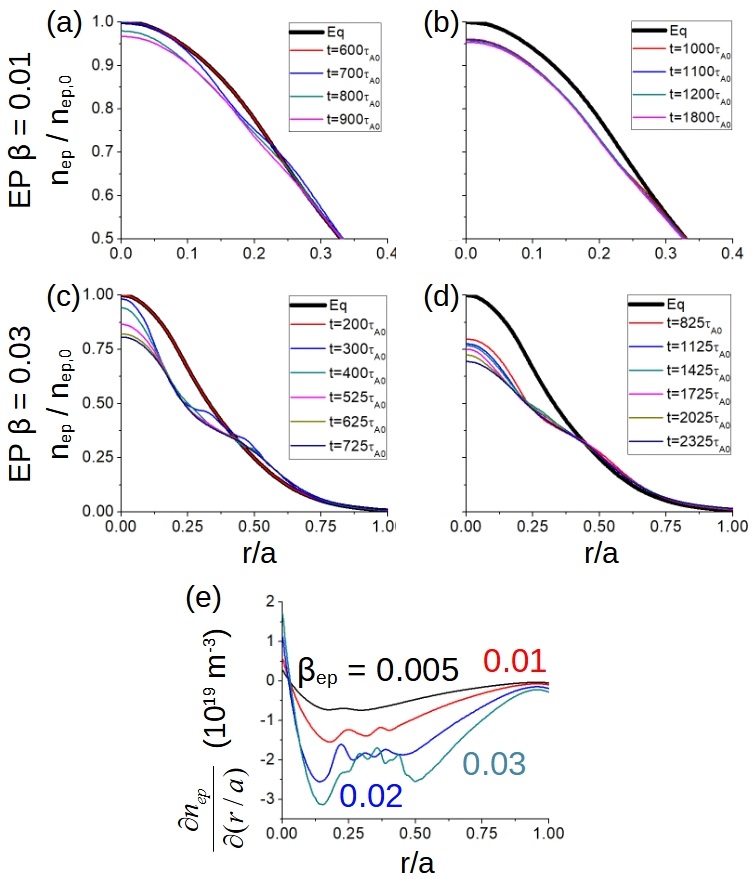}
\caption{EP density profile evolution. Simulation with EP $\beta = 0.01$ during (a) the linear and beginning of the saturation phase as well as (b) the late saturation phase. Simulation with EP $\beta = 0.03$ during (c) the linear and saturation phase as well as (d) the bursting phase. Critical gradient $\gamma_{c} = \partial n_{ep} / \partial (r/a)$ in the simulations with EP $\beta = 0.005$ (black line), $0.01$ (red), $0.02$ (blue) and $0.03$ (cyan).}\label{FIG:16}
\end{figure}

Stiff EP density profiles were measured above a given threshold of the NBI injection intensity in DIII-D discharges \cite{94}, also observed in the saturated phase of the nonlinear simulation in the soft MHD limit, for example in the EP $\beta=0.01$ case from $t = 900 \tau_{A0}$ (figure~\ref{FIG:16}, panels a and b). On the other hand, the simulations in the hard MHD limit show an evolving EP profile both in the saturated and bursting phases, for example in the EP $\beta=0.03$ case (figure~\ref{FIG:16} panels c and d). This is explained by the combined effect of zonal structures and nonlinear energy transfers between different toroidal mode families on the AE stability. This also affects the induced EP transport. Consequently, a non-stiff EP density profile is a consequence of an incompletely saturated AE. Thus, the discrepancy with the experiment can be explained by the reduced number of modes included in the simulations. We speculate that a system with larger number of overlapping AEs may lead to the hard MHD limit transition for a lower EP $\beta$. This analysis will be the topic of a future study dedicated to compare nonlinear simulations and DIII-D discharges if the NBI operates in the hard MHD limit. Nevertheless, the present analysis indicates the role of nonlinear effects is essential to understand the triggering of the burst events and may have an impact on the EP density profile stiffness. Panel (e) shows the critical gradient ($\gamma_{c} = \partial n_{ep} / \partial (r/a)$) calculated in simulations with different EP $\beta$ values. The analysis indicates $\gamma_{c}$ increases between the simulations in the soft MHD limit. On the other hand, $\gamma_{c}$ is similar between $r/a = 0.25$-$0.5$, the radial location where the AEs overlap, for the simulations in the hard MHD limit. Consequently, the simulation results are consistent with the critical-gradient behavior observed in DIII-D experiments \cite{95}. Nevertheless, the study results are not conclusive for the hard MHD limit because the simulations are not fully saturated. The simulations also indicate that the critical-gradient formalism requires the thermal plasma to reach a dynamic equilibrium with the EP drive. Thus, the thermal plasma profiles must not show large excursions with respect to the equilibrium profiles so that a stable balance between zonal structures regulating the AE amplitude and the EP transport induced by the AEs is maintained. If the burst activity is too strong, with recurrent large amplitude burst events and large excursions of the plasma profiles, then the critical-gradient behavior could break down because EP and thermal plasma profiles are not stiff. In other words, a second transition in the hard MHD limit can be induced if the NBI injection intensity is large enough, leading to different system dynamics and EP transport.

\begin{figure}[h!]
\centering
\includegraphics[width=0.5\textwidth]{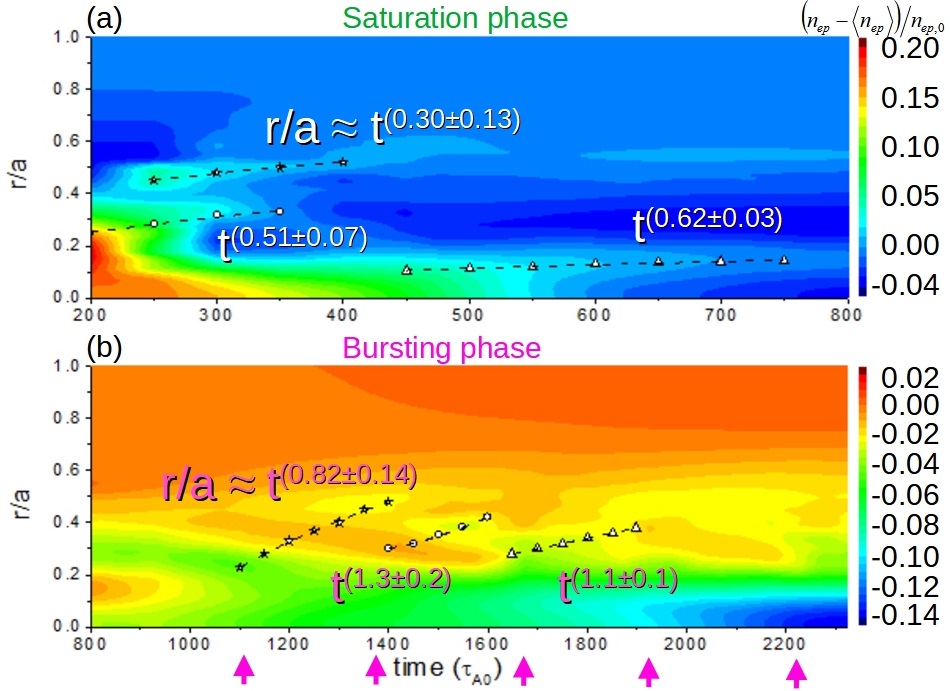}
\caption{Deviation of the normalized EP density with respect to the time averaged EP density in the simulation with EP $\beta = 0.03$. (a) Saturation phase and (b) bursting phase. Yellow stars indicate the outwards displacement of EP density fluid elements induced by unstable AEs in the saturation phase. Stars, circles and triangles tracks the outwards displacement of EP density fluid elements caused by different AEs. Dashed black lines indicate the result of the regression $r/a = A + B*t^{C}$ for each .}\label{FIG:17}
\end{figure}

Figure~\ref{FIG:17} shows the deviation of the normalized EP density with respect to the time averaged EP density ($\langle n_{ep} \rangle$), defined as $\overline{\rm n}_{ep} = (n_{ep} - \langle n_{ep} \rangle) / n_{ep,0}$. The aim of this analysis is to measure the time scaling of the outward propagation rate of the $\overline{\rm n}_{ep}$ contours as a proxy of the EP transport. The study is performed for the simulation with EP $\beta = 0.03$ during the saturation, panel (a), and bursting, panel (b), phases. The analysis indicates the outwards propagation of EP density fluid elements induced by unstable AEs. Particularly, the EP transport observed during the saturation phase around $r/a=0.25$ is caused by the $n=3$ AE (circles), around $0.5$ by the $n=6$ AE (stars) and around $0.1$ by the unstable $0/0$ and $0/1$ modes (triangles). On the other hand, the EP transport induced during the bursting phase is caused by the burst events triggered between $r/a = 0.25$-$0.3$, the radial location where $n=3$ and $6$ AEs overlap. The EP transport linked to the burst $E1$ is indicated by stars, $E2$ by circles and $E3$ by triangles. The features of the EP transport can be analyzed by fitting the radial location of the EP density fluid element with respect to the simulation time. The regression calculated is $r/a = A + B*t^{C}$, indicated by black dashed lines. The regression exponent in the saturation phase is close to $1/2$, that is to say, the EP transport is mainly diffusive. On the other hand, the fit exponents calculated during the bursts are close to unity, implying a ballistic EP transport that may reveal the triggering of avalanche-like events, consistent with DIII-D observations. It should be noted that the simulations results are also consistent with the avalanche-like events observed in NSTX and the massive migration of energetic ions in JT60U plasma, as well as the numerical analysis performed which suggested that the no linear mode interaction may cause the onset of EP avalanches \cite{96,97}.

The nonlinear analysis indicates an improved AE stability and a lower EP drive if the NBI voltage is reduced, both in the soft and hard MHD limits. The amplitude of the AEs is reduced as well as the EP transport. It should be noted that the simulations show that a reduction of the NBI voltage is more efficient to weaken the EP transport and improve the plasma heating efficient than decreasing the NBI power if the NBI operates in the soft MHD limit. On the other hand, these optimization trends are unclear if the NBI operates in the hard MHD limit because the EP transport is dominated by the burst activity induced by the radial overlapping of multiple AEs. The conclusions obtained from the simulations show similar trends for the AE stability with respect to DIII-D experimental observations that demonstrated a lower AE activity if the NBI operates with a lower voltage for a fixed power injection \cite{46}.

The analysis of the NBI radial deposition on the AE saturation phase indicates off-axis NBI injection  reduces the EP drive and the amplitude of the unstable AEs, although the unstable plasma region is more radially extended compared to an on-axis injection. Nevertheless, the EP transport obtained in the on-axis case is similar to the off-axis case if the NBI is deposited at $r/a \leq 0.15$. On the other hand, the EP transport is $40 \%$ lower if the NBI is deposited at $r/a = 0.3$. That means an off-axis NBI injection may improve the plasma heating efficiency if the deposition region is at $r/a \geq 0.15$. The weakening of the AE activity in DIII-D plasma, particularly RSAE and core TAEs was observed experimentally if the NBI is injected off-axis \cite{45,98}. The simulations also show a reduction of the reverse shear region depth in the inner plasma region for the on-axis NBI case, caused by a decrease of the safety factor near the magnetic axis due to zonal currents generated by the AEs. On the other hand, the depth of the reverse shear region increases if the NBI is deposited off axis.

The effect of the collisionality is not directly included in the study. Collisionality is implicitly included through the diffusivities that are present in each dynamical equation since diffusive transport is caused by collisional effects. The collisionality regulates the scattering of the EPs pitch angle, modifying the EP drive as well as the AE stability and saturation phase properties \cite{99,100,101}. The simulations performed are dedicated to an analysis of the destabilizing effect of the passing EP component, thus the effect of the EP pitch angle scattering caused by the collisionality is not directly included. Consequently, any effect on the AE stability caused by the variation of the collisionality along the normalized minor radius or the thermal plasma evolution are not included in the model. This model limitation may cause a deviation between simulations prediction and observations because the EP drive and AE stability / saturation are not perfectly reproduced. Nevertheless, the range of parameters scanned may minimize the impact of this limitation by reproducing robust AE saturation phases in the soft and hard MHD regimes, neglecting the collisionality effect may not modify the global conclusion of the manuscript, just alter the EP $\beta$ required for the transition between soft and hard MHD regimes.

The study of the linear AE stability using the evolved equilibrium profiles calculated in the nonlinear simulations opens up the possibility of comparing the output of linear and nonlinear simulations; this can be useful to demonstrate the use of linear simulations to identify optimization trends with respect to the NBI operational regime. The analysis indicates the discrepancy between linear and nonlinear simulations increases as the EP drive enhances, leading to a rather large deviation for the case of an NBI operating in the hard MHD limit. On the other hand, the prediction of linear and nonlinear simulations is fairly similar if the NBI operates in the soft MHD limit. This is explained by an enhancement of the EP transport induced by radially overlapped AEs in the hard MHD limit, thus linear simulations lack critical information about the EP density and thermal plasma profiles evolution, further perturbed as the EP drive increases. Consequently, the application of linear simulations to identify optimization trends with respect to the NBI operational regime is justified if the NBI only causes a local perturbation of the plasma. It should be noted that future devices as ITER may show a large radial overlapping between different EP resonances \cite{102,103,104,105}, thus the application of linear models to study the AE stability may be limited to NBI operational regimes with rather low power injection leading to a plasma in the soft MHD limit. 

\ack
This work was supported by the Comunidad de Madrid under the project 2019-T1/AMB-13648. The authors want to thank W.W. Heidbrink for useful discussion. Two of the authors (H.B. and D.Z.) have received financial support from the AIM4EP project (ANR-21-CE30-0018), funded by the French National Research Agency (ANR).

\hfill \break

\end{document}